\documentclass[]{aa}

\usepackage{graphicx}
\usepackage{txfonts}
\usepackage{amsmath}
\usepackage{adjustbox}
\usepackage{hyperref}
\hypersetup{colorlinks=true,allcolors=[rgb]{0,0,0.5}}
\usepackage[dvipsnames]{xcolor}
\usepackage[version=4]{mhchem}
\usepackage{soul}
\usepackage{placeins}

\makeatletter
\renewcommand*\aa@pageof{, page \thepage{} of \pageref*{LastPage}}
\makeatother

\begin{document}





      \title{Why heterogeneous cloud particles matter:}
      \subtitle{Iron-bearing species and cloud particle morphology affects exoplanet transmission spectra}

   \author{S. Kiefer \inst{1, 2, 3}
            \and
            D. Samra \inst{2}
            \and
            D. A. Lewis \inst{2}
            \and
            A. D. Schneider \inst{1, 4}
            \and
            M. Min \inst{5}
            \and
            L. Carone \inst{2}
            \and
            L. Decin \inst{1}
            \and
            Ch. Helling \inst{2, 3}
            }

   \institute{Institute of Astronomy, KU Leuven, Celestijnenlaan 200D, 3001 Leuven, Belgium\\
              \email{sven.kiefer@kuleuven.be}
         \and
              Space Research Institute, Austrian Academy of Sciences, Schmiedlstrasse 6, A-8042 Graz, Austria
         \and
             Institute for Theoretical Physics and Computational Physics, Graz University of Technology, Petersgasse 16 8010 Graz
         \and
              Centre for ExoLife Sciences, Niels Bohr Institute, {\O}ster Voldgade 5, 1350 Copenhagen, Denmark
        \and
             SRON Netherlands Institute for Space Research, Leiden, The Netherlands
             }

   \date{Received ...; accepted ...}

  \abstract
   { The possibility of observing spectral features in exoplanet atmospheres with space missions like the \textit{James Webb Space Telescope} (JWST) and \textit{Atmospheric Remote-sensing Infrared Exoplanet Large-survey} (ARIEL) necessitates the accurate modelling of cloud particle opacities. In exoplanet atmospheres, cloud particles can be made from multiple materials and be considerably chemically heterogeneous. Therefore, assumptions on the morphology of cloud particles are required to calculate their opacities.
   }
   { The aim of this work is to analyse how different approaches to calculate the opacities of heterogeneous cloud particles affect cloud particle optical properties and how this may effect the interpretation of data observed by JWST and future missions.
   }
   { We calculate cloud particle optical properties using seven different mixing treatments: four effective medium theories (EMTs: Bruggeman, Landau-Lifshitz-Looyenga (LLL), Maxwell-Garnett, and Linear), core-shell, and two homogeneous cloud particle approximations. We conduct a parameter study using two-component materials to study the mixing behaviour of 21 commonly considered cloud particle materials for exoplanets. To analyse the impact on observations, we study the transmission spectra of HATS-6b, WASP-39b, WASP-76b, and WASP-107b.
   }
   { Materials with large refractive indices, like iron-bearing species or carbon, can change the optical properties of cloud particles when they comprise less than 1\% of the total particle volume. The mixing treatment of heterogeneous cloud particles also has an observable effect on transmission spectroscopy. Assuming core-shell or homogeneous cloud particles results in less muting of molecular features and retains the cloud spectral features of the individual cloud particle materials. The predicted transit depth for core-shell and homogeneous cloud particle materials are similar for all planets used in this work. If EMTs are used, cloud spectral features are broader and cloud spectral features of the individual cloud particle materials are not retained. Using LLL leads to less molecular features in transmission spectra compared to Bruggeman.
   }
  {}


   \maketitle
%

\section{Introduction}
\label{sec:Introduction}

    Cloud formation models for exoplanet atmospheres predict significant cloud coverage for all but the very hottest exoplanets \citep{roman_clouds_2021, helling_exoplanet_2023}. However, until recently, the only observable impact of clouds on observations were characteristically flat transmission spectra \citep{bean_ground-based_2010, kreidberg_clouds_2014, espinoza_access_2019, spyratos_transmission_2021, libby-roberts_featureless_2022}. Only with observations from the \textit{James Webb Space Telescope} (JWST) is it now possible to directly measure spectral features characteristic of, for example, silicate clouds in exoplanet atmospheres \citep{grant_jwst-tst_2023, dyrek_so2_2023}.

    \begin{table*}
        \centering
        \caption{Assumptions made by exoplanet atmosphere studies.}
        \label{tab:summary}
        \begin{tabular}{l l l l l l}
            \hline\hline
                Paper                           & Planet     & Code        & Use case & Approach    & Cloud Species     \\
             \hline \hline
                \citet{dyrek_so2_2023}          & WASP-107b  & ARCiS       & Transm.  & Brg.      & MgSiO$_3$, SiO$_2$, SiO        \\
                                                &            & pRT         & Transm.  & par.      & -                              \\
                                                &            & pRT         & Transm.  & hom.      & MgSiO$_3$, SiO$_2$, KCl        \\
                \citet{grant_jwst-tst_2023}     & WASP-17b   & ATMO        & Transm.  & grey      & -                              \\
                                                &            & VIRGA       & Transm.  & hom.      & SiO$_2$, Al$_2$O$_3$           \\
                                                &            & POSEIDON    & Transm.  & grey      & -                              \\
                                                &            & POSEIDON    & Transm.  & hom.      & SiO$_2$                        \\
                                                &            & pRT         & Transm.  & hom.      & SiO$_2$                        \\
                \citet{alderson_early_2023}     & WASP-39b   & VIRGA       & Transm.  & hom.      & MnS, Na$_2$S, MgSiO$_3$        \\
                                                &            & ATMO        & Transm.  & grey      & -                              \\
                                                &            & PHOENIX     & Transm.  & grey      & -                              \\
                \citet{rustamkulov_early_2023}  & WASP-39b   & PICASO 3.0  & Transm.  & grey      & -                              \\
                                                &            & ScCHIMERA   & Transm.  & grey      & -                              \\
                                                &            & ATMO        & Transm.  & grey      & -                              \\
                                                &            & PHOENIX     & Transm.  & grey      & -                              \\
                \citet{ahrer_early_2023}        & WASP-39b   & VIRGA       & Transm.  & hom.      & MnS, Na$_2$S, MgSiO$_3$        \\
                \citet{feinstein_early_2023}    & WASP-39b   & ATMO        & Transm.  & par.      & -                              \\
                                                &            & PHOENIX     & Transm.  & par.      & -                              \\
                                                &            & VIRGA       & Transm.  & hom.      & MnS, Na$_2$S, MgSiO$_3$        \\
                                                &            & ScCHIMERA   & Transm.  & grey      & -                              \\
                                                &            & ScCHIMERA   & Transm.  & hom.      & MgSiO$_3$                      \\
                \citet{arfaux_coupling_2024}    & WASP-39b   & "unnamed"   & Transm.  & hom.      & Na$_2$S, MgSiO$_3$             \\
                \citet{lustig-yaeger_jwst_2023} & LHS 457 b  & CHIMERA     & Transm.  & grey      & -                              \\
                \citet{min_arcis_2020}          & Multiple   & ARCiS       & Transm.  & Brg.      & SiO$_2$, Fe, FeS, Al$_2$O$_3$, C, \\
                                                &            &             &          &           & SiC, TiO$_2$, VO, MgSiO$_3$    \\
                \citet{wong_optical_2020}       & HAT-P-12b  & CARMA       & Transm.  & cs.       & TiO$_2$, Al$_2$O$_3$, Mg$_2$SiO$_4$, \\
                                                &            &             &          &           & MnS, Na2S, KCl, ZnS            \\
                \citet{powell_transit_2019}     & Test case  & CARMA       & Transm.  & hom.      & TiO$_2$, Fe, Mg$_2$SiO$_4$, Al$_2$O$_3$ \\
                \citet{kempton_reflective_2023} & GJ 1214b   & HyDRo       & Therm.   & hom.      & KCl                            \\
                \citet{chubb_exoplanet_2022}    & WASP-43b   & ARCiS       & Therm.   & par.      & -                              \\
                \citet{gao_universal_2021}      & Grid       & CARMA       & Therm.   & cs.       & TiO$_2$, Fe, Cr, KCl,  \\
                                                &            &             &          &           & Mg$_2$SiO$_4$, MnS,     \\
                                                &            &             &          &           & Al$_2$O$_3$, Na$_2$S \\
                \citet{webber_effect_2015}      & Kepler-7b  & "unnamed"   & Therm.   & hom.      & MgSiO$_3$, Mg$_2$SiO$_4$, Fe   \\
                \citet{demory_inference_2013}   & Kepler-7b  & "unnamed"   & Therm.   & hom.      & Mg$_2$SiO$_4$                  \\
                \citet{lee_modelling_2023}      & HAT-P-1b   & Mini-Cloud  & GCM      & LLL       & TiO$_2$, Al$_2$O$_3$, Fe, Mg$_2$SiO$_4$ \\
                \citet{christie_impact_2022}    & GJ 1214b   & UM          & GCM      & hom.      & KCl, ZnS                       \\
                \citet{komacek_patchy_2022}     & Grid       & MITgcm      & GCM      & cs.       & TiO$_2$, Fe, Cr, KCl, \\
                                                &            &             &          &           & Mg$_2$SiO$_4$, Cr, MnS, \\
                                                &            &             &          &           & Al$_2$O$_3$, Na$_2$S \\
                \citet{roman_clouds_2021}       & Grid       & RM-GCM      & GCM      & hom.      & KCl, ZnS, Na$_2$S, MnS, \\
                                                &            &             &          &           & SiO$_2$, Mg$_2$SiO$_4$, VO,  \\
                                                &            &             &          &           & Ca$_2$SiO$_4$, CaTiO$_2$, Al$_2$O$_3$,  \\
                                                &            &             &          &           & Fe, Cr$_2$O$_3$, Ni \\
                \citet{tan_global_atmospheric_2021} & Test case & MITgcm   & GCM      & hom.      & MgSiO$_3$                       \\
                \citet{lines_simulating_2018}   & HD209458b  & UM          & GCM      & Brg./LLL  & TiO$_2$, SiO, SiO$_2$,         \\
                                                &            &             &          &           & MgSiO$_3$, Mg$_2$SiO$_4$       \\
                \citet{lee_dynamic_2016}        & HD189733b  & "unnamed"   & GCM      & Brg./LLL  & TiO$_2$, SiO, SiO$_2$,         \\
                                                &            &             &          &           & Mg$_2$SiO$_4$, MgSiO$_3$       \\
             \hline
        \end{tabular}
        \tablefoot{Studies that consider absorption features within transmission spectra are labelled "Transm.". Studies that consider absorption features within the thermal emission of the planet are labelled "Therm.". Studies that consider radiative feedback of clouds within global circulation models are labelled "GCM". All cloud species are in the solid phase ([s]). Abbreviations used in this table: Bruggeman (Brg.), core-shell (cs.), homogeneous cloud particles (hom.), grey cloud deck (grey), and parameterised cloud description (par.).}
    \end{table*}

    Many different cloud models for exoplanet atmospheres exist in literature which use different assumptions in order to describe the ensemble of particles that make up the clouds. The formation of clouds in exoplanet atmospheres starts with the formation of cloud condensation nuclei (CCNs) onto which other cloud particle materials can grow. Species like TiO$_2$ \citep{sindel_revisiting_2022}, SiO \citep{lee_dust_2015}, VO \citep{lecoq-molinos_vanadium_2024} and KCl \citep{gao_microphysics_2018} are predicted to nucleate and grow homogeneously. Other materials, for example MgSiO$_3$ or ZnS, form on the surface of the CCNs. This leads to heterogeneous cloud particles. In the cloud formation model of \citet{arfaux_coupling_2024}, hazes made from soot particles act as CCNs onto which MgSiO$_3$ and Na$_2$S can form. To calculate the cloud particle opacities, they assume that the soot particles are neglectable and cloud particles are considered to be homogeneous. The cloud formation model CARMA \citep{turco_one-dimensional_1979, toon_multidimensional_1988, jacobson_modeling_1994, ackerman_model_1995, bardeen_numerical_2008, gao_microphysics_2018} starts with TiO$_2$ and KCl nucleation onto which other species can grow \citep[e.g. Fe, Cr, Al$_2$O$_3$, Mg$_2$SiO$_4$, Cr, MnS, and Na$_2$S;][]{komacek_patchy_2022}. Their cloud particles are therefore made from one core species (either TiO$_2$ or KCl) and one shell species. This cloud particle morphology is called core-shell or core-mantle. The cloud formation model of \citet{helling_dust_2006} \citep[see also][]{helling_dust_2001, woitke_dust_2003, woitke_dust_2004, helling_dust_2004} starts with TiO$_2$, SiO, KCl, NaCl, and C nucleation onto which multiple other species can grow. This model considers surface reactions for the growth of cloud particle materials that are not stable as gas-phase molecules. They assume that all cloud particle materials are well-mixed.

    Radiative transfer calculation for exoplanet atmospheres require the calculation of cloud opacities. In computationally expensive retrieval frameworks, clouds are often considered as a grey cloud deck where the opacity is assumed wavelength independent \citep[e.g.][]{alderson_early_2023, rustamkulov_early_2023} or are wavelength-dependently parameterised \citep[e.g.][]{chubb_exoplanet_2022, feinstein_early_2023}. However, cloud particles have spectral features in transmission spectra, in particular for wavelengths above 8~$\mu$m \citep{wakeford_transmission_2015}. While grey clouds allow to account for the general muting of molecular features in transmission spectra, it has been shown that non-grey clouds are needed to accurately interpret observations \citep{powell_transit_2019, feinstein_early_2023}. Many studies therefore consider separate populations of homogeneous cloud particles to account for cloud particle features \citep[e.g.][]{demory_inference_2013, powell_transit_2019, roman_clouds_2021, tan_global_atmospheric_2021, christie_impact_2022, grant_jwst-tst_2023, feinstein_early_2023, kempton_reflective_2023, arfaux_coupling_2024}. Other models also consider the optical properties of heterogeneous cloud particles \citep{lee_dynamic_2016, lines_simulating_2018, komacek_patchy_2022, min_arcis_2020, komacek_patchy_2022, dyrek_so2_2023, lee_dynamically_2024}. A summary of currently used assumptions for cloud opacity calculations can be found in Table~\ref{tab:summary}.

    The optical properties of homogeneous materials has been widely studied \citep[see e.g. Table~2 of the online material\footnote{Available at https://zenodo.org/records/13373168.} or][]{kitzmann_optical_2018}. The optical properties of heterogeneous cloud particles, however, are more complicated to derive. If the cloud particle materials are well-mixed within the cloud particle, effective medium theories\footnote{Also called effective medium approximations.} (EMTs) can be used (see Sect.~\ref{sec:theory_emt}). The three most commonly used EMTs are Bruggeman \citep{bruggeman_berechnung_1935}, Landau-Lifshitz-Looyenga \citep[LLL;][]{landau_electrodynamics_1960, looyenga_dielectric_1965}, and Maxwell-Garnett \citep{garnett_xii_1904}. While comparison studies between EMTs exist, they were mostly done for materials common in solid state physics \citep[see e.g.][]{kolokolova_scattering_2001, du_use_2004, franta_comparison_2005}.

    Modelling cloud particles as spherical is a common assumption for studies of exoplanet atmospheres \citep[e.g.][]{kitzmann_optical_2018, mai_exploring_2019, sanghavi_cloudy_2021, komacek_patchy_2022, arfaux_coupling_2024, grant_jwst-tst_2023, jaiswal_scattering_2023, lee_modelling_2023}.
    Although efforts have been made to model the opacities of fractal, or otherwise irregularly shaped, cloud particles in exoplanet atmospheres \citep[e.g.][]{ohno_clouds_2020}. However, calculating more accurate refractive indices at the highest level of theory, requires the full knowledge of the material distribution within the grain \citep[see e.g.][]{lodge_aerosols_2023}. For example with the Discrete Dipole approximation \citep{draine_discrete-dipole_1994}, or the T-Matrix Method \citep{mackowski_calculation_1996}. Other theories have sought to simplify the interactions of irregularly shaped particles, either by assuming a statistical distribution of simpler shapes such as ellipsoids \citep{bohren_absorption_1983, min_scattering_2003}, or `hollow spheres' \citep{min_scattering_2003, min_modeling_2005, samra_mineral_2020}. In addition to this, approximations to the electromagnetic field as it interacts with an irregularly shaped grain, to allow for self-interaction of the cloud  particle, have also been tried \citep{tazaki_light_2016, tazaki_light_2018}. Lastly, the shape of cloud particles, and their material distribution can have important consequences for the polarisation of cloud-scattered light \citep{draine_sensitivity_2024,chubb_modelling_2024}.

    The opacities of heterogeneous particles is studied for aerosols in Earth's atmosphere \cite[e.g.][]{mishchenko_applicability_2016, stegmann_regional_2017}. \citet{chylek_scattering_1988} found that the predictions from Bruggeman and Maxwell-Garnett for water inclusions in acrylic produce similar, but not identical, results to experiments. However, this only holds for well-mixed particles. \citet{liu_inhomogeneity_2014} have shown that EMTs only give accurate results if the characteristic size of individual inclusions is less than 0.12 times the wavelength they interact with. Several studies have looked at core-shell particles \citep[e.g.][]{katrib_products_2004, lee_organic_2020}. \citet{mcgrory_mie_2022} performed experiments using silica aerosols within a mist of sulphuric acid particles. They showed that aerosols within their setup had a core-shell morphology with silica being the core and sulphuric acid building the shell. The existence of core-shell morphology within Earth's aerosols is also supported by observations from \citet{unga_microscopic_2018}. Using remote sensing, they found that 60\% of urban and 20\% of desert aerosols present residuals of a core-shell morphology.

    In this paper, we study the optical properties of heterogeneous cloud particles in exoplanet atmospheres. We consider four EMTs (Bruggeman, LLL, Maxwell-Garnett, and Linear), the core-shell morphology, and two homogeneous cloud particle approximations (Sect.~\ref{sec:theory}). We conduct a parameter study using two-component materials (Sect.~\ref{sec:2comp}) to study the mixing behaviour of 21 commonly considered cloud particle materials for exoplanets \citep[see e.g.][]{powell_formation_2018, helling_sparkling_2019, gao_aerosols_2021}: Fe[s], FeO[s], Fe$_2$O$_3$[s], Fe$_2$SiO$_4$[s], FeS[s], TiO$_2$[s], SiO[s], CaTiO$_3$[s], SiO$_2$[s], MgO[s], MgSiO$_3$[s], Mg$_2$SiO$_4$[s], Al$_2$O$_3$[s],  NaCl[s], KCl[s], C[s], C$_\mathrm{amorphous}$[s], ZnS[s], Na$_2$S[s], MnO[s], MnS[s]. To study how heterogeneous cloud particle affects potential observations, we analyse the pressure dependent cloud structure and the transmission spectra of HATS-6b, WASP-39b, WASP-76b, and WASP-107b (Sect.~\ref{sec:planets}). The results are discussed in Sect.~\ref{sec:dis} and the conclusion is given in Sect.~\ref{sec:conclusion}.

\section{Theoretical Basis}
\label{sec:theory}

    Cloud particles are an important opacity source in exoplanet atmospheres with complex optical properties. Sect.~\ref{sec:theory_cloudop} describes how cloud particle absorption and scattering coefficients can be calculated. In this work we assume spherical cloud particles which allows one to use Mie theory to calculate their absorption and scattering efficiency (Sect.~\ref{sec:theory_mie}). Cloud particles in exoplanet atmospheres can be made from a mixture of different cloud particle materials \citep[see e.g.][]{helling_sparkling_2019, gao_universal_2021}. The morphology of cloud particles in exoplanet atmospheres is currently unknown and depends on how the cloud particles are formed. A common assumption is that all materials are well-mixed \citep[see e.g.][]{lee_dynamic_2016, lines_simulating_2018, min_arcis_2020, helling_cloud_2021}. This assumption allows one to use EMTs to calculate the effective refractive index of heterogeneous cloud particles (Sect.~\ref{sec:theory_emt}). Another assumption is the core-shell morphology which occurs when condensates form a shell around a CCN core \citep[see e.g.][]{gao_aerosols_2021, komacek_patchy_2022}. The core-shell morphology and two additional non-mixed treatments are considered in this study (Sect.~\ref{sec:theory_bas}).

    \subsection{Cloud particle opacities}
    \label{sec:theory_cloudop}

    The amount of radiation that gets absorbed or scattered by cloud particles is determined by their absorption coefficient $\kappa_\mathrm{abs}^\mathrm{cloud} (\lambda)$~[cm$^2$~kg$^{-1}$] and the scattering coefficient $\kappa_\mathrm{sca}^\mathrm{cloud} (\lambda)$~[cm$^2$~kg$^{-1}$]. Assuming spherical particles with radii $a$~[cm], these coefficients are given by:
    \begin{align}
        \label{eq:theory_kappa_abs}
        \kappa_\mathrm{abs}^\mathrm{cloud}(\lambda) &= \int_{a_\mathrm{min}}^\infty \frac{\pi a^2 f_\mathrm{d} (a)}{\rho_\mathrm{gas}} ~Q_\mathrm{abs} (a, \lambda, \epsilon_\mathrm{eff}) ~da \\
        \label{eq:theory_kappa_sca}
        \kappa_\mathrm{sca}^\mathrm{cloud}(\lambda) &= \int_{a_\mathrm{min}}^\infty \frac{\pi a^2 f_\mathrm{d} (a) }{\rho_\mathrm{gas}} ~Q_\mathrm{sca} (a, \lambda, \epsilon_\mathrm{eff}) ~ (1 - g) ~da
    \end{align}
    where $f_\mathrm{d}(a)$~[cm$^{-4}$] is the cloud particle size distribution, $a_\mathrm{min}$~[cm] the smallest radius of a particle for it to be considered a cloud particle, $\rho_\mathrm{gas}$~[g~cm$^{-3}$] is the gas density, and $\lambda$~[cm] the wavelength of the radiation. The variable $\epsilon_\mathrm{eff}$ is the effective refractive index which is further discussed in Sect.~\ref{sec:theory_emt}. The variables $Q_\mathrm{abs}$, $Q_\mathrm{sca}$, and $g$ are the absorption efficiency, scattering efficiency, and anisotropy factor, respectively. These three variables are calculated using Mie Theory (Sect.~\ref{sec:theory_mie}).

    \subsection{Mie Theory}
    \label{sec:theory_mie}

    Mie theory \citep{mie_beitrage_1908} describes the solution to the Maxwell equations for the interaction of radiation with a sphere. The solution depends on the wavelength of the radiation $\lambda$, the radius of the sphere $a$, the dielectric constant $\epsilon_\mathrm{s}$ of the sphere, and the dielectric constant of the surrounding medium $\epsilon_\mathrm{m}$. For cloud particles, the surrounding medium is vacuum which has $\epsilon_\mathrm{m} = 1$. Mie theory calculations only depend on the relative size of the cloud particle and the wavelength of the radiation. Thus Mie calculations use the size parameter:
    \begin{align}
        x = 2 \pi \frac{ a }{\lambda}
    \end{align}
    A detailed description on how the absorption and scattering coefficients are calculated for cloud particles can be found in \citet{gail_physics_2013}.

    \subsection{Effective medium theory}
    \label{sec:theory_emt}

    To calculate the interaction of cloud particles and radiation, the refractive index $n + ik$ of the cloud particles must be known. The refractive index is directly related to the dielectric constant $\epsilon$ of a material:
    \begin{align}
        \epsilon = (n + ik)^2
    \end{align}
    The real part of the refractive index $n$ describes the ratio of the speed of light and the phase velocity of light in the medium. The imaginary part $k$ describes the attenuation of the electromagnetic wave travelling through the medium.

    Many studies of the optical properties of homogeneous cloud particle materials in exoplanet atmospheres exist \citep[e.g.][]{wakeford_transmission_2015, kitzmann_optical_2018, potapov_importance_2022}. The data used in this work can be found in Table~2 of the online material\footnote{Available at https://zenodo.org/records/13373168.}. Mixed materials on the other hand have complicated dielectric properties which depend on their composition and material distribution. Hence, simplifying assumptions have to be made to investigate these properties for clouds in exoplanet atmospheres. If one assumes a heterogeneous material to be well-mixed, its optical properties can be approximated by the effective dielectric constant
    \begin{align}
        \epsilon_\mathrm{eff} = (n_\mathrm{eff} + ik_\mathrm{eff})^2
    \end{align}
    EMTs solve the electric fields for a mixture of materials to derive $\epsilon_\mathrm{eff}$. While $\epsilon$ is a precise description of the properties of a homogeneous material, $\epsilon_\mathrm{eff}$ is an approximation under the assumption that a well-mixed material can be described equivalently to a homogeneous material.

    The Maxwell-Garnett approximation \citep{garnett_xii_1904, markel_introduction_2016} was derived to explain colours in metal glasses and in metallic films. It assumes one dominant material with small inclusions. This simplifies the calculation of the effective refractive index by neglecting interactions between inclusions. The effective dielectric constant is then given by the following formula:
    \begin{align}
        \label{eq:maxwell-garnett}
        \frac{\epsilon_\mathrm{eff} - \epsilon_d}{\epsilon_\mathrm{eff} + 2\epsilon_d} = \sum_i f_i \frac{\epsilon_i - \epsilon_d}{\epsilon_i + 2\epsilon_d}
    \end{align}
    where $\epsilon_d$ is the dielectric constant of the dominant material, $\epsilon_i$ are the dielectric constants of inclusions, and $f_i = V_i / V_\mathrm{tot}$ is the volume fractions of the inclusions of material $i$, $V_i$~[cm$^{3}$] the volume of material $i$ within a cloud particle, and $V_\mathrm{tot}$~[cm$^{3}$] is the total volume of the cloud particle. For this work, we assume the cloud particle material with the largest volume fraction to be the dominant material and all others to be inclusions. Because this theory was derived for small inclusions, the accuracy of Maxwell-Garnett becomes questionable for volume fractions of inclusions above 10$^{-6}$ \citep{belyaev_electrodynamic_2018}. In exoplanet atmospheres, the volume fraction of the second most dominant cloud particle material is often predicted to be over 1\% \citep{helling_exoplanet_2023}. Therefore, Maxwell-Garnett might not always be applicable.

    The LLL approximation \citep{landau_electrodynamics_1960, looyenga_dielectric_1965} was derived for dispersive, powder-like mixtures. The derivation of the effective dielectric constant assumes small variations of the dielectric constant within small spherical inclusions of a larger sphere. The effective dielectric constant is then given by:
    \begin{align}
        \label{eq:lll}
        \epsilon_\mathrm{eff} = \left( \sum_i f_i \sqrt[3]{\epsilon_i} \right)^3 .
    \end{align}
    This equation presents an analytical solution for the effective dielectric constant $\epsilon_\mathrm{eff}$. Due to the short computational time, this technique is of special interest for larger frameworks which include heterogeneous cloud particles, like global circulation models \citep[GCMs; see e.g.,][]{lee_modelling_2023}.

    The Bruggeman approximation \citep{bruggeman_berechnung_1935} assumes small, homogeneous inclusions. The derivation of the effective dielectric constant is done by replacing a small spherical inclusion within a sphere with a different material. The effective dielectric constant is then derived using the following equation:
    \begin{align}
        \label{eq:bruggeman}
        \sum_i f_i \frac{\epsilon_i - \epsilon_\mathrm{eff}}{\epsilon_i + 2\epsilon_\mathrm{eff}} = 0 .
    \end{align}
    The calculation of $\epsilon_\mathrm{eff}$ using Bruggeman requires a minimisation scheme making it computationally intensive. This is in particular impractical for the implementation of cloud particles in larger frameworks. To combat this problem, \citet{lee_dynamic_2016} used a Newton-Raphson minimisation, but fall back to LLL if no solution can be found within a given number of iteration steps.

    The most simplistic approximation of the effective dielectric constants is derived by linearly summing dielectric constants for each material, weighted by their volume fraction \citep{mackwell_comparative_2014, kahnert_modelling_2015}:
    \begin{align}
        \label{eq:linear}
        \epsilon_\mathrm{eff} = \sum_i f_i \epsilon_i .
    \end{align}
    This approximation is called Linear from here on. It is important to note that the Linear approximation is not derived from the electrical properties of the mixed material like the other EMTs. Thus, it is unlikely to accurately represent the optical properties of a mixed material. However, this technique can be used to, for example, finding a suitable starting condition for the minimisation of Bruggeman.

    \subsection{Non-mixed treatments of cloud particles}
    \label{sec:theory_bas}

    \begin{figure}
        \centering
        \includegraphics[width=\hsize]{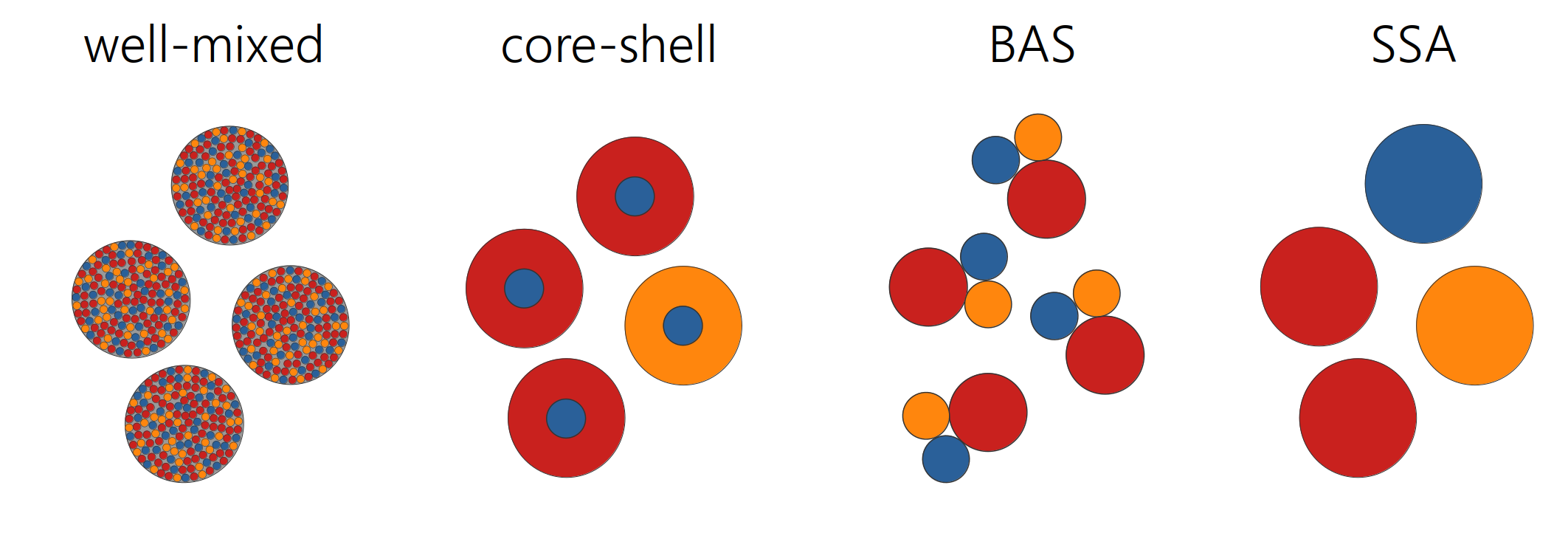}
        \caption{Representation of the cloud particle morphology assumptions behind well-mixed, core-shell, BAS, and SSA cloud particles.}
        \label{fig:visual_emts}
    \end{figure}

    In the most general case, cloud particle opacities will depend on their exact composition, shape, and material distribution. EMTs simplify this complexity by assuming well-mixed grains. However, certain materials within a cloud particle might form larger homogeneous inclusions making EMTs no longer applicable. In this case, non-mixed treatments are required. A visualisation of the different theories used in this work is shown in Fig.~\ref{fig:visual_emts}.

    The core-shell morphology describes a particle consisting of a homogeneous core surrounded by a homogeneous shell. This morphology is consistent with cloud particle materials forming a shell around a CCN core and is assumed by, for example, CARMA. Typically, both core and shell are assumed spherical. The absorption and scattering efficiency for a core-shell morphology can be calculated similar to Mie theory \citep{toon_algorithms_1981}. In the case where multiple species can condense simultaneously, each shell material is assumed to form their own population of cloud particles. Therefore, each cloud particle consists at most of two cloud particle materials.

    The Batch Approximation of Spheres (BAS) assumes that each cloud particle material forms a single sphere within the cloud particle. The volume of this sphere is equal to the volume fraction of the cloud particle material. Each cloud particle therefore consists of multiple spheres, each made from a single material. To simplify the calculation of the cloud particle opacities, interactions between the spheres and their overlaps are neglected. The absorption and scattering coefficients (see Eq.~\ref{eq:theory_kappa_abs} and \ref{eq:theory_kappa_sca}) are then adjusted accordingly:
    \begin{align}
        \label{eq:theory_bas_abs}
        \kappa^\mathrm{cloud}_\mathrm{abs}(\lambda) &=  \int_{a_\mathrm{min}}^\infty \frac{\pi f_\mathrm{d} (a)}{\rho_\mathrm{gas}}  \sum_{i \in \mathcal{M}} \left( f_i^{1/3} a \right)^2 Q_\mathrm{abs} (f_i^{1/3} a, \lambda, \epsilon_i) ~da\\
        \label{eq:theory_bas_sca}
        \kappa^\mathrm{cloud}_\mathrm{sca}(\lambda) &= \int_{a_\mathrm{min}}^\infty  \frac{\pi f_\mathrm{d} (a)}{\rho_\mathrm{gas}} \sum_{i \in \mathcal{M}} \left( f_i^{1/3} a \right)^2 Q_\mathrm{sca} (f_i^{1/3} a, \lambda, \epsilon_i) (1 - g) ~da
    \end{align}
    where $\mathcal{M}$ is the set of all cloud particle materials. Since overlapping is neglected, this represents an upper limit of the cloud particle opacity. Furthermore, the interaction between cloud particle materials could lead to spectral features which cannot be reproduced with BAS.

    The System of Spheres Approximation (SSA) assumes that all cloud particle materials are separated into homogeneous, spherical cloud particles with a radius of $\langle a \rangle$. This assumption is equivalent to other studies which assume multiple homogeneous species with the same cloud particle radius but different number densities \citep[e.g.][]{roman_clouds_2021, dyrek_so2_2023}. For SSA, the cloud particle number density of each cloud particle material is proportional to the volume fraction $f_i$. The absorption and scattering coefficients (see Eq.~\ref{eq:theory_kappa_abs} and \ref{eq:theory_kappa_sca}) are than adjusted accordingly:
    \begin{align}
        \label{eq:theory_ssa_abs}
        \kappa^\mathrm{cloud}_\mathrm{abs} (\lambda) &= \int_{a_\mathrm{min}}^\infty \frac{\pi a^2}{\rho_\mathrm{gas}} \sum_{i \in \mathcal{M}} f_i~f_\mathrm{d} (a)~ Q_\mathrm{abs} (a, \lambda, \epsilon_i)~da \\
        \label{eq:theory_ssa_sca}
        \kappa^\mathrm{cloud}_\mathrm{sca} (\lambda) &= \int_{a_\mathrm{min}}^\infty \frac{\pi a^2}{\rho_\mathrm{gas}} \sum_{i \in \mathcal{M}} f_i~f_\mathrm{d} (a)~ Q_\mathrm{sca} (a, \lambda, \epsilon_i) (1 - g)~da
    \end{align}
    While SSA does not require any additional assumption on the optical properties of the cloud particles, it is inconsistent with cloud formation models which predict that materials like Mg$_2$SiO$_4$[s] form heterogeneously \citep[see e.g.][]{helling_dust_2006, gao_microphysics_2018}.

\section{Approach}

    After examining the theoretical basis, we continue to detail our approach. The calculation of the cloud particle opacities are explained in Sect.~\ref{sec:approch_opac} with the transmission spectra calculations following in Sect.~\ref{sec:theory_trans}. All methods laid out here are available within the software package \texttt{Claus}.

    \subsection{Cloud opacity calculation}
    \label{sec:approch_opac}

    For all calculations in this study, we assume spherical cloud particles and calculate the absorption and scattering efficiencies using Mie theory. However, the solution of Mie theory includes an infinite sum and thus has to be solved numerically. A comparison between different Mie-solvers can be found in Appendix~\ref{app:sec:mietheory}. Overall, we found little differences between the implementations and thus decided to use \texttt{PyMieScatt} for this study \citep{sumlin_retrieving_2018}. Within \texttt{Claus}, \texttt{miepython} \citep{wiscombe_mie_1979, prahl_miepython_2023}, and \texttt{Miex} \citep{wolf_mie_2004} are also available.

    The cloud particle opacities are calculated by approximating the optical properties of all cloud particle sizes with cloud particles of an average radius $\langle a \rangle$~[cm]. While size distributions do affect the optical properties of cloud layers, a detailed investigation of their impact on transmission spectra goes beyond the scope of this paper. To calculate the cloud particle number density, we use the cloud particle mass fraction ${\rho_\mathrm{v}}/{\rho_\mathrm{gas}}$:
    \begin{align}
        n_d = \frac{\rho_\mathrm{v}}{\rho_\mathrm{gas}} \frac{1}{\frac{3}{4} \pi \langle a \rangle^3 \rho_c}
    \end{align}
    where $\rho_\mathrm{v}$~[g~cm$^{-3}$] is the cloud particle mass per atmospheric volume, $\rho_\mathrm{gas}$~[g~cm$^{-3}$] is the gas density, and $\rho_c$~[g~cm$^{-3}$] is the cloud particle material density. The cloud particle absorption coefficient from Eq.~\ref{eq:theory_kappa_abs} is thus given by:
    \begin{align}
        \kappa_\mathrm{abs}^\mathrm{cloud}(\lambda) &= \frac{\pi \langle a \rangle^2}{\frac{4}{3} \pi \langle a \rangle^3 \rho_c} \frac{\rho_\mathrm{v}}{\rho_\mathrm{gas}}  ~Q_\mathrm{abs} (\langle a \rangle, \lambda, \epsilon_\mathrm{eff})
    \end{align}
    The opacity calculations from Eq. \ref{eq:theory_kappa_sca}, \ref{eq:theory_bas_abs}, \ref{eq:theory_bas_sca}, \ref{eq:theory_ssa_abs}, and \ref{eq:theory_ssa_sca} are adjusted accordingly. The cloud particle opacities therefore depend on the following cloud particle properties:

    \begin{itemize}
        \item The average cloud particle radius $\langle a \rangle$~[cm]
        \item The cloud particle material density $\rho_c$~[g~cm$^{-3}$]
        \item The cloud mass fraction $\rho_\mathrm{v} / \rho_\mathrm{gas}$
        \item The refractive index of the cloud particle $\epsilon_\mathrm{eff}$ or $\epsilon_\mathrm{i}$
    \end{itemize}

    To study the opacities of heterogeneous cloud particles, we first analyse two-component materials (Sect.~\ref{sec:2comp}). A two-component material \{A,~B\} is a composite material made from the mixture of the materials A and B. They find a broad use in solid state physics and hence multiple effective medium studies exist \citep[see e.g.][]{du_use_2004, ghanbarian_permeability_2016}. In this study, we use materials commonly considered in exoplanet atmospheres and vary the relative abundance of A and B. First, the two-component material \{Fe[s],~Mg$_2$SiO$_4$[s]\} is used to compare the effective refractive index from different EMTs (Sect.~\ref{sec:2comp_emts}). Afterwards, we consider the mixture of Mg$_2$SiO$_4$[s] with 16 other commonly considered cloud particle materials\footnote{For hot Jupiters we consider Fe[s], FeO[s], Fe$_2$O$_3$[s], Fe$_2$SiO$_4$[s], FeS[s], TiO$_2$[s], SiO[s], CaTiO$_3$[s], SiO$_2$[s], MgO[s], MgSiO$_3$[s], Mg$_2$SiO$_4$[s], Al$_2$O$_3$[s],  NaCl[s], KCl[s], C[s], and C$_\mathrm{amorphous}$[s].} in hot Jupiters (Sect.~\ref{sec:2comp_hot}) and ZnS[s] with 5 commonly considered cloud particle materials\footnote{For temperate atmospheres, we consider NaCl[s], KCl[s], ZnS[s], Na$_2$S[s], MnO[s], and MnS[s].} of temperate atmospheres (Sect.~\ref{sec:2comp_temp}). To compare the absorption and scattering coefficients of EMTs and non-mixed approaches, we use again the two-component material \{Fe[s],~Mg$_2$SiO$_4$[s]\} (Sect.~\ref{sec:2comp_csvsemt}).

    In hot Jupiter atmospheres, cloud particles can consist of more than two materials. To study the optical properties of considerably heterogeneous cloud particles, we use the results from a detailed micro-physical cloud model \citep{helling_dust_2006}. First, we investigate the optical properties of clouds at various heights within the atmosphere of WASP-39b (Sect.~\ref{sec:planets_hats6b}). Afterwards, we analyse the impact of cloud particle mixing treatments on the transmission spectrum of WASP-39b and HATS-6b (Sect.~\ref{sec:planets_trans}). The best target for studying the composition and morphology of cloud particles is an exoplanet where cloud materials have been detected in the gas phase and whose atmosphere is predicted to contain clouds. One such planet is WASP-76b, which we are analysing in more detail (Sect.~\ref{sec:planets_wasp76b}). Since the optical properties of cloud particles do not only affect the predictions of forward models but also retrievals, we take a closer look at WASP-107b using the results from the ARCiS retrieval of \citet{dyrek_so2_2023} (Sect.~\ref{sec:planets_wasp107b}).

    \subsection{Cloud modelling}

    In Sect.~\ref{sec:planets}, the cloud structure of 4 planets is studied: HATS-6b, WASP-39b, WASP-76b, and WASP-107b. The cloud modelling approaches for these planets is described in this section.

    The cloud structures of HATS-6b , WASP-39b, and WASP-76b are done using a hierarchical forward modelling approach. The temperature and wind structure are modelled using expeRT/MITgcm \citep{carone_equatorial_2020, schneider_exploring_2022} which is cloud-free. The GCM simulations were done by \citet{carone_wasp-39b_2023} for WASP-39b, \citet{kiefer_under_2024} for HATS-6b, and \citet{schneider_no_2022} for WASP-76b. The output of the GCM is then used to calculate the cloud structure. For this, one dimensional temperature, pressure, and vertical velocity profiles were extracted from the GCM and used as input for the kinetic cloud model developed by \citet{helling_dust_2006} \citep[see also][]{woitke_dust_2003, woitke_dust_2004, helling_dust_2004, helling_dust_2008, helling_sparkling_2019}. This model includes micro-physical nucleation, bulk growth, and evaporation which are combined with gravitational settling, element consumption, and replenishment. A list of all nucleating species and cloud particle materials considered can be found in Table~1 of the online material\footnote{Available at https://zenodo.org/records/13373168.}. The cloud structures were derived by \citet{carone_wasp-39b_2023} for WASP-39b and \citet{kiefer_under_2024} for HATS-6b. For WASP-76b, the cloud structures are produced here for the first time. All 1D profiles used in this work are shown in Fig.~6 (HATS-6b), Fig.~7 (WASP-39b), and Fig.~8 (WASP-76b) of the online material\footnote{Available at https://zenodo.org/records/13373168.}.

    To calculate the transmission spectra of all three planets, 1D profiles from multiple latitudes within the morning and evening terminator were considered. The opacities of the cloud particles were calculated based on these profiles and the different mixing treatments as described in Sect.~\ref{sec:theory}. For the calculation of core-shell opacities, the dominant nucleation species was selected as the core species. For WASP-39b and HATS-6b, this is SiO[s]. For WASP-76b this is TiO$_2$[s].

    The cloud structure of WASP-107b is taken from the ARCiS \citep{min_arcis_2020} retrieval results of \citet{dyrek_so2_2023}. In contrast to the other planets, the temperature-pressure profile, gas-phase abundances, and cloud particle properties ($V_\mathrm{s}/V_\mathrm{tot}$, $\langle a \rangle$, $n_d$) are derived from a free retrieval based on JWST MIRI observations. This means that the model does not include a micro-physical cloud formation description. The opacities of the cloud particles were calculated based on the retrieved cloud particle properties and the different mixing treatments as described in Sect.~\ref{sec:theory}. For the calculation of core-shell opacities, SiO[s] was selected as the core species.

    \subsection{Transmission spectrum}
    \label{sec:theory_trans}

    To calculate the transmission spectrum of cloudy exoplanet atmospheres, the cloud particle opacities are added as an additional opacity source to petitRADTRANS \citep{molliere_petitradtrans_2019, molliere_retrieving_2020, alei_large_2022}. The number densities of gas-phase species are calculated within the cloud model, thereby assuming chemical equilibrium and including the depletion of gas-phase species due to cloud formation\footnote{The only exception to this is the transmission spectrum of WASP-107b which uses the retrieval results instead.}.

    The following species were considered as gas-phase opacity species within the radiative transfer \citep{chubb_exomol_2020}:
    H$_2$O \citep{polyansky_exomol_2018},
    CO$_2$ \citep{yurchenko_exomol_2020},
    CO \citep{li_rovibrational_2015},
    TiO \citep{mckemmish_exomol_2019},
    Na \citep{piskunov_vald_1995},
    and K \citep{piskunov_vald_1995}.
    Rayleigh scattering is included by considering
    H$_2$ \citep{dalgarno_rayleigh_1962}
    and He \citep{chan_refractive_1965}.
    Collision-induced absorption (CIA) is considered from H$_2$--H$_2$ and H$_2$--He \citep{borysow_collision-induced_1988, borysow_collision-induced_1989-1, borysow_collision-induced_1989, borysow_high-temperature_2001, richard_new_2012, borysow_collision-induced_2002}.

    The references for the optical properties of homogeneous materials are listed in Table~2 of the online material\footnote{Available at https://zenodo.org/records/13373168.}. Because no opacity data is available for CaSiO$_3$[s], its refractive index is set to vacuum for all calculations\footnote{Another approach is to approximate the opacity of CaSiO$_3$[s] with MgSiO$_3$[s]. A comparison between the two approaches can be found in Appendix~\ref{sec:app_casio3}.}. Heterogeneous particles are calculated using EMT (see Sect.~\ref{sec:theory_emt}). Planetary parameters used for the transmission spectrum calculation can be found in Table~1 of the online material\footnote{Available at https://zenodo.org/records/13373168.}. Generally, it is expected that clouds mute molecular features in the optical and near infrared, resulting in a flat transmission spectrum \citep{bean_ground-based_2010, kreidberg_clouds_2014, powell_transit_2019}. Signatures of metal-oxide bonds of cloud particle materials can impact the spectra typically around and above 10~$\mu$m \citep{wakeford_transmission_2015, grant_jwst-tst_2023, dyrek_so2_2023}.

\section{Mixing behaviour of cloud particle materials}
    \label{sec:2comp}

    We present the effective refractive index of mixed cloud particles. Here, we focus on two-component materials in order to understand the differences between the mixing treatments and different cloud particle materials. In Sect~\ref{sec:2comp_emts}, we compare the predicted effective refractive index and absorption efficiency from LLL, Bruggeman, Maxwell-Garnett, and Linear. In Sect.~\ref{sec:2comp_hot}, we analyse materials commonly predicted to form clouds in hot Jupiter atmospheres and in Sect.~\ref{sec:2comp_temp}, materials commonly predicted for temperate exoplanets. Non-mixed treatments do not assume an effective refractive index but employ different types of Mie theory calculation. In Sect.~\ref{sec:2comp_csvsemt}, we compare the absorption and scattering efficiencies from LLL, core-shell, BAS, and SSA.

    \subsection{Effective refractive index from EMTs}
    \label{sec:2comp_emts}

    \begin{figure}
        \centering
        \includegraphics[width=\hsize]{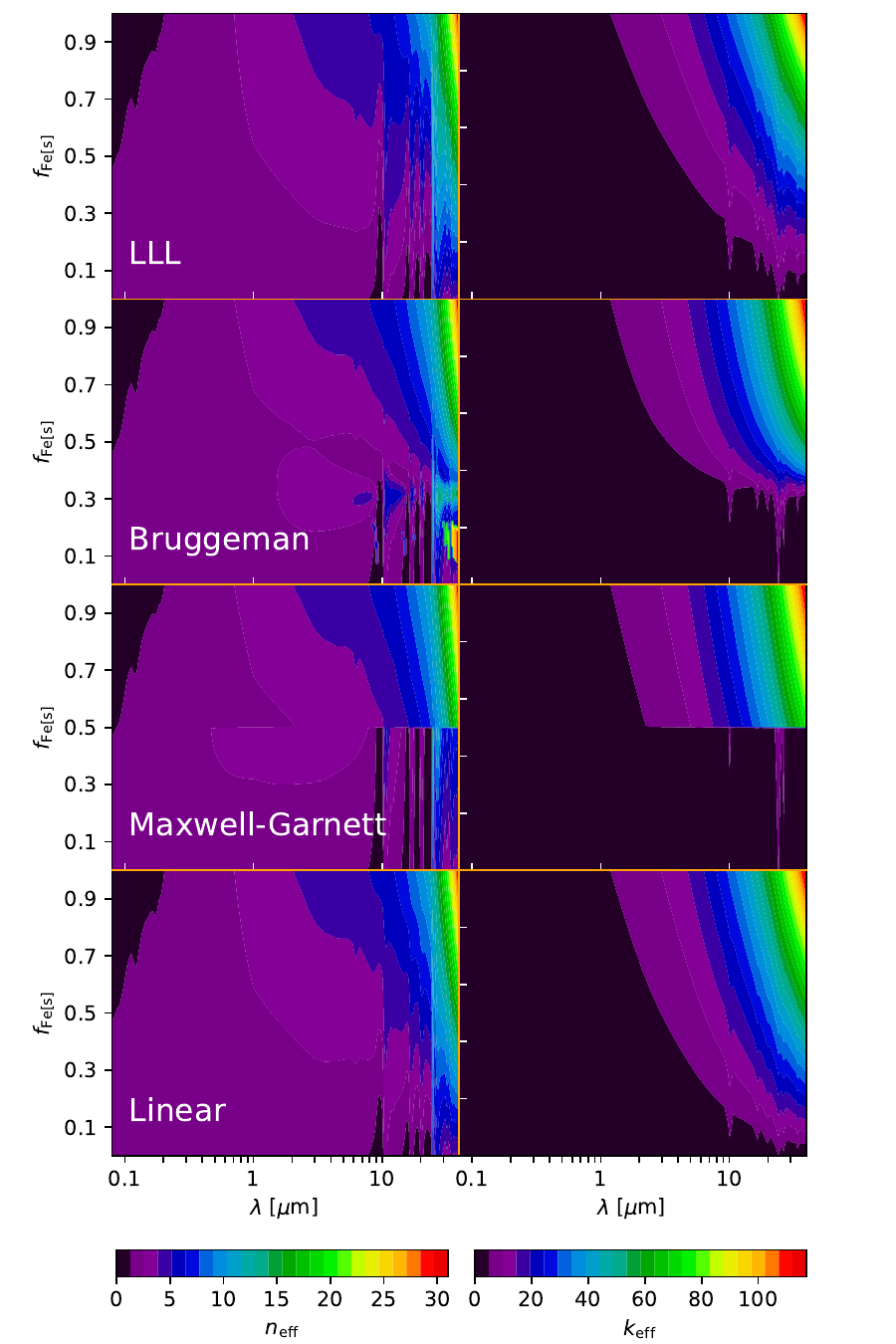}
        \caption{Effective refractive index of the two-component material \{Fe[s],~Mg$_2$SiO$_4$[s]\} calculated using different EMTs. \textbf{Left:} real part $n_\mathrm{eff}$. \textbf{Right:} imaginary part $k_\mathrm{eff}$.}
        \label{fig:mat_emt}
    \end{figure}

    \begin{figure}
        \centering
        \includegraphics[width=\hsize]{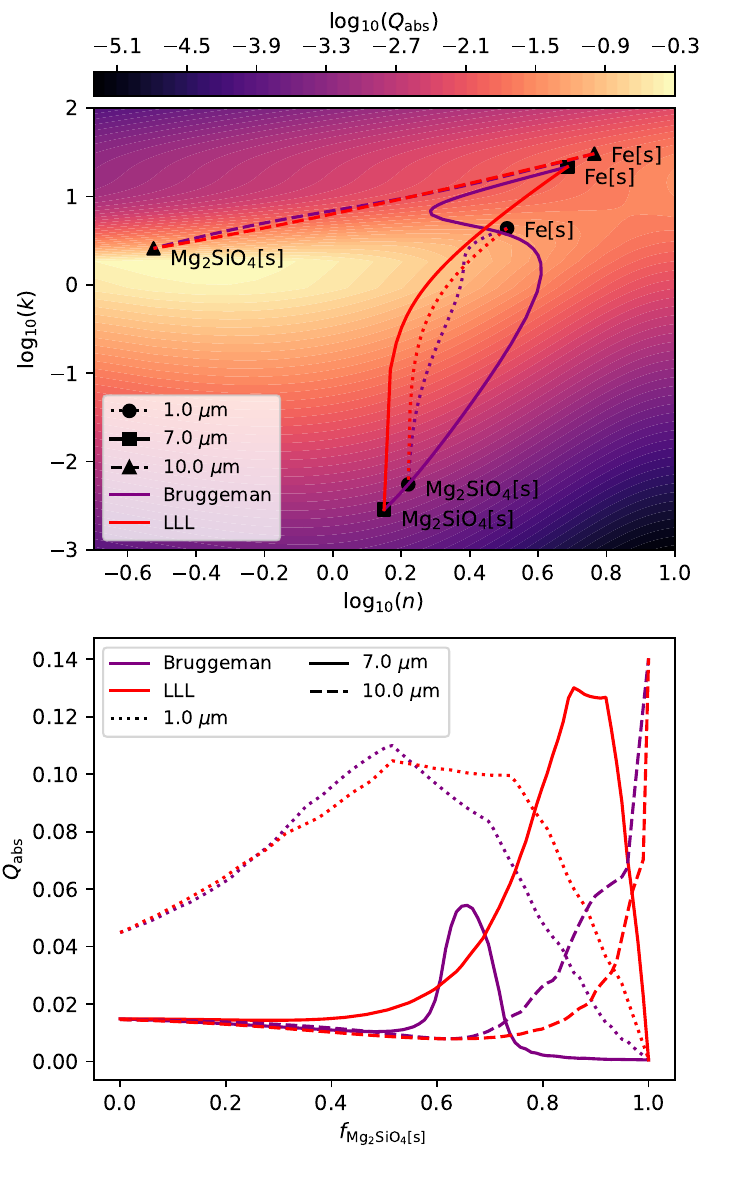}
        \caption{\textbf{Top:} Absorption efficiency $Q_\mathrm{abs}$ for a range of $n$ and $k$ values at $x = 0.1$. Also shown are the $k_\mathrm{eff}$ and $n_\mathrm{eff}$ values for the two-component material \{Fe[s],~Mg$_2$SiO$_4$[s]\} at wavelengths 1~$\mu$m, 7~$\mu$m, and 10~$\mu$m. The end points denote homogeneous materials. The lines between the end points represent different volume fractions of the two materials. \textbf{Bottom:} Absorption efficiency $Q_\mathrm{abs}$ for different volume fractions of the two-component material \{Fe[s],~Mg$_2$SiO$_4$[s]\} at wavelengths 1~$\mu$m, 7~$\mu$m, and 10~$\mu$m.}
        \label{fig:iron_explor}
    \end{figure}

    The effective refractive index is calculated using Bruggeman (Eq.~\ref{eq:bruggeman}), LLL (Eq.~\ref{eq:lll}), Maxwell-Garnett (Eq.~\ref{eq:maxwell-garnett}), and Linear (Eq.~\ref{eq:linear}). These calculation depend only on the refractive index and the volume mixing ratio of the cloud particle material. Here, we use the example two-component material \{Fe[s],~Mg$_2$SiO$_4$[s]\}. Since forsterite (Mg$_2$SiO$_4$[s]) is often discussed as a dominant cloud particle material \citep{powell_formation_2018, gao_aerosol_2020, helling_cloud_2021, helling_exoplanet_2023}, we chose it to be the first component of all two-component materials in this section. Iron-bearing species are of great interest for this study because of their large imaginary part $k$ of the refractive index compared to other cloud particle materials. We therefore choose Fe[s] as the second material. The effective refractive index of \{Fe[s],~Mg$_2$SiO$_4$[s]\} for a wavelength range of 0.08~$\mu$m to 39~$\mu$m and for Fe[s] volume fractions between 0 and 1 can be seen in Fig.~\ref{fig:mat_emt}. The absolute relative differences between the EMTs can be seen in Fig.~1 of the online material\footnote{Available at https://zenodo.org/records/13373168.}.

    In Fig..~\ref{fig:mat_emt}, LLL and Linear predict similar effective refractive indices for most wavelengths. In contrast to the other EMTs, both LLL and Linear show a strong increase in $k_\mathrm{eff}$ with increasing Fe[s] volume fraction. From 10~$\mu$m to 30~$\mu$m, features in the $n_\mathrm{eff}$ and $k_\mathrm{eff}$ values can be seen which are caused by peaks in the values of the refractive index of Mg$_2$SiO$_4$[s]. These features diminish with increasing Fe[s] volume fraction. The biggest difference between LLL and Linear is around 10~$\mu$m where LLL predicts up to 50\% smaller $n_\mathrm{eff}$ and $k_\mathrm{eff}$ values. This wavelength corresponds to a vibrational resonance of Si-O bonds \citep[see e.g.][]{gunde_vibrational_2000, sogawa_infrared_2006}. Therefore, this difference can be explained by the third root of the dielectric constant within LLL (see Eq.~\ref{eq:lll}) which reduces the impact of resonances on the effective dielectric constant.

    Maxwell-Garnett shows an abrupt change in $k_\mathrm{eff}$ and $n_\mathrm{eff}$ values at 0.5 volume fraction. This corresponds to the change in the dominant cloud particle material. When Mg$_2$SiO$_4$[s] is the dominant cloud particle material the values for $k_\mathrm{eff}$ and $n_\mathrm{eff}$ are lower for wavelengths above 8~$\mu$m. Maxwell-Garnett showing differences to other EMTs for larger volume fractions of iron is not unexpected since the validity of Maxwell-Garnett is only guaranteed for volume fractions below 10$^{-6}$ \citep{belyaev_electrodynamic_2018}. However, our results show that for Mg$_2$SiO$_4$[s] volume fractions of less than 0.1, Maxwell-Garnett and Bruggeman differ by less than 10\% for this particular two-component material.

    Bruggeman and LLL predict different effective refractive indices for all wavelengths above 0.2~$\mu$m. This finding is in agreement with other studies which also found significant differences between Bruggeman and LLL \citep{du_use_2004}. Especially for Fe[s] volume fraction below 0.3 and for wavelengths above 1~$\mu$m, Bruggeman predicts significantly lower $k_\mathrm{eff}$. The computational cost of LLL, Maxwell-Garnett and Linear is roughly the same. Linear is $\sim$10\% faster than LLL, and LLL is $\sim$5\% faster than Maxwell-Garnett. Bruggeman, however, takes roughly 1000 times longer than LLL. This is mainly due to the minimisation required to solve Bruggeman whereas LLL, Maxwell-Garnett and Linear have analytical solutions.

    The effective refractive index calculation for the two component material \{Fe[s],~Mg$_2$SiO$_4$[s]\} clearly highlights the limitations of Maxwell-Garnett and Linear for applications to heterogeneous cloud particles in exoplanet atmospheres. From here on, we will therefore focus on the EMTs Bruggeman and LLL.
    
    The effective refractive index directly impacts the absorption efficiency $Q_\mathrm{abs}$ and scattering efficiency $Q_\mathrm{sca}$. To investigate how Mie theory and EMTs interact, we calculate $Q_\mathrm{abs}$ for the Bruggeman, Maxwell-Garnett, and LLL EMTs. We use the two-component material \{Fe[s],~Mg$_2$SiO$_4$[s]\} and assume a size parameter of $x = 0.1$ for the Mie calculation. The top panel of Fig.~\ref{fig:iron_explor} shows the values of $Q_\mathrm{abs}$ for different $k_\mathrm{eff}$ and $n_\mathrm{eff}$ values. Within this figure, the change of the effective refractive indices with volume fraction is shown for the wavelengths 1~$\mu$m, 7~$\mu$m, and 10~$\mu$m. The end points of each line denote homogeneous materials. The bottom panel of Fig.~\ref{fig:iron_explor} shows the $Q_\mathrm{abs}$ along the volume fraction lines shown in the top panel. The same evaluation for a size parameter of $x = 1$, the extinction efficiency, the scattering efficiency, and the two component material \{TiO$_2$[s],~Mg$_2$SiO$_4$[s]\}, is shown in Fig.~2 of the online material\footnote{Available at https://zenodo.org/records/13373168.}.

    For 1~$\mu$m, both LLL and Bruggeman predict a higher $Q_\mathrm{abs}$ for volume fractions around 0.5 than a homogeneous particle made from either Fe[s] or Mg$_2$SiO$_4$[s] alone. This can be explained by a maxima of Q$_\mathrm{abs}$ for $k$ around $\sim$1.4 as predicted by Mie-theory. \citet{min_modeling_2005} showed that this maximum is a consequence of the spherical particles assumption. Because $k_\mathrm{eff}$ and $n_\mathrm{eff}$ values for a volume fraction around 0.5 are closer to the maxima, these particles have a larger $Q_\mathrm{abs}$ value. At 7~$\mu$m a similar behaviour as for 1~$\mu$m can be seen. However, LLL approaches the maxima much closer than Bruggeman, resulting in a peak in $Q_\mathrm{abs}$ values around 0.9 Mg$_2$SiO$_4$[s] volume fraction. Bruggeman, on the other hand, avoids the maxima and only peaks around 0.65 Mg$_2$SiO$_4$[s] volume fraction. At 10~$\mu$m, the refractive index of homogeneous Mg$_2$SiO$_4$[s] particles is very close to the maxima, resulting in large $Q_\mathrm{abs}$ values. The higher the volume fraction of Fe[s], the weaker this peak in opacity becomes. Here, all EMTs predict a similar behaviour with volume fraction.

    \subsection{Effective refractive index for clouds in hot Jupiter}
    \label{sec:2comp_hot}

    \begin{figure}
        \centering
        \includegraphics[width=\hsize]{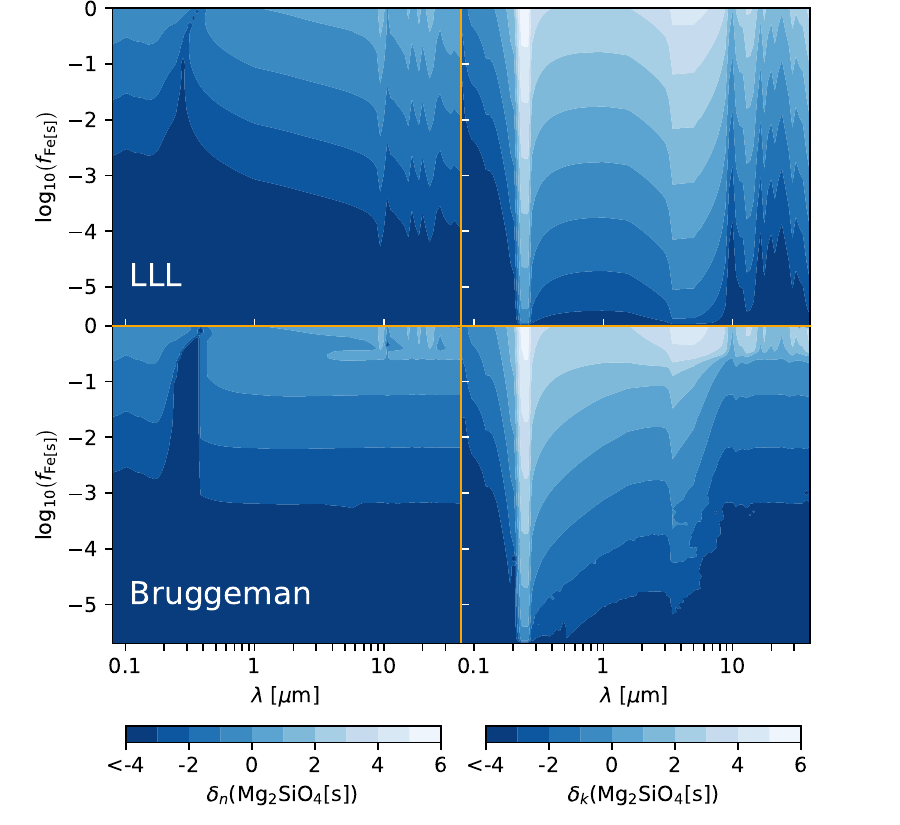}
        \caption{Differences of the real and imaginary part of the refractive index (Eq.~\ref{eq:diff}) of the two-component material \{Mg$_2$SiO$_4$[s],~Fe[s]\} compared to the refractive index of Mg$_2$SiO$_4$[s].}
        \label{fig:fe_mat}
    \end{figure}

    \begin{figure*}[hbtp!]
        \centering
        \includegraphics[width=0.49\hsize]{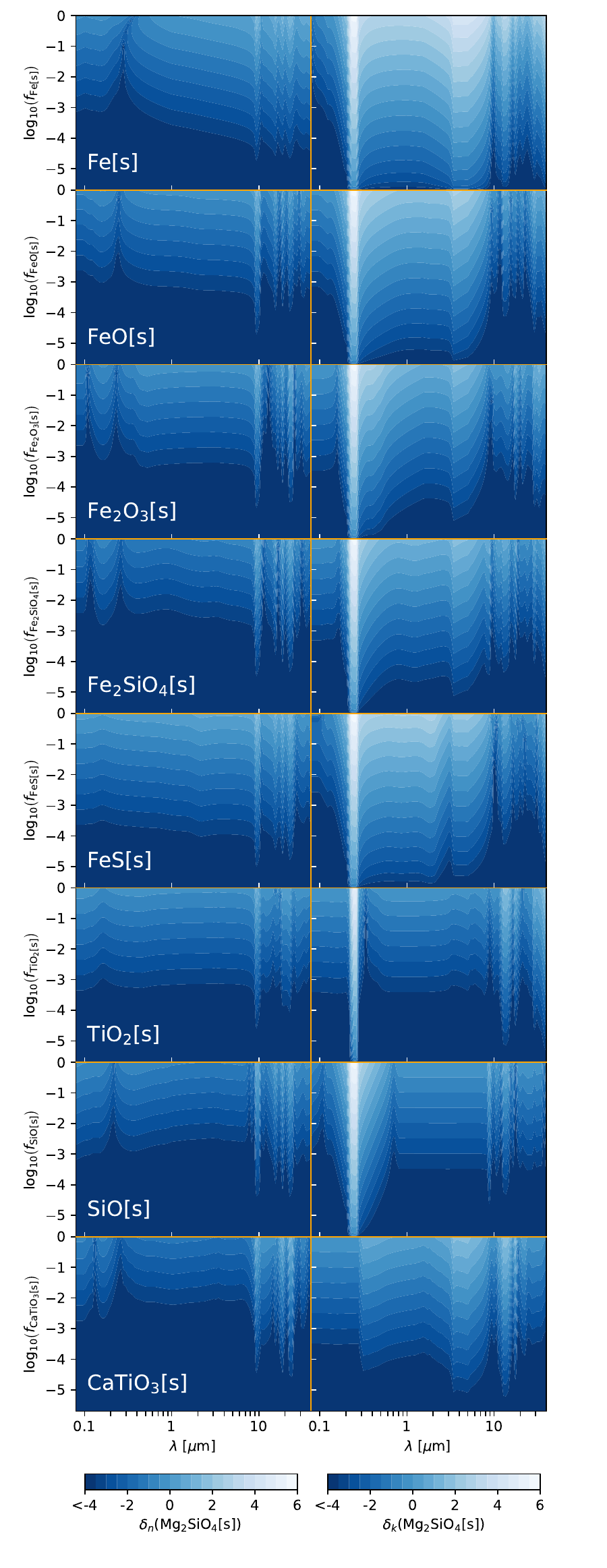}
        \includegraphics[width=0.49\hsize]{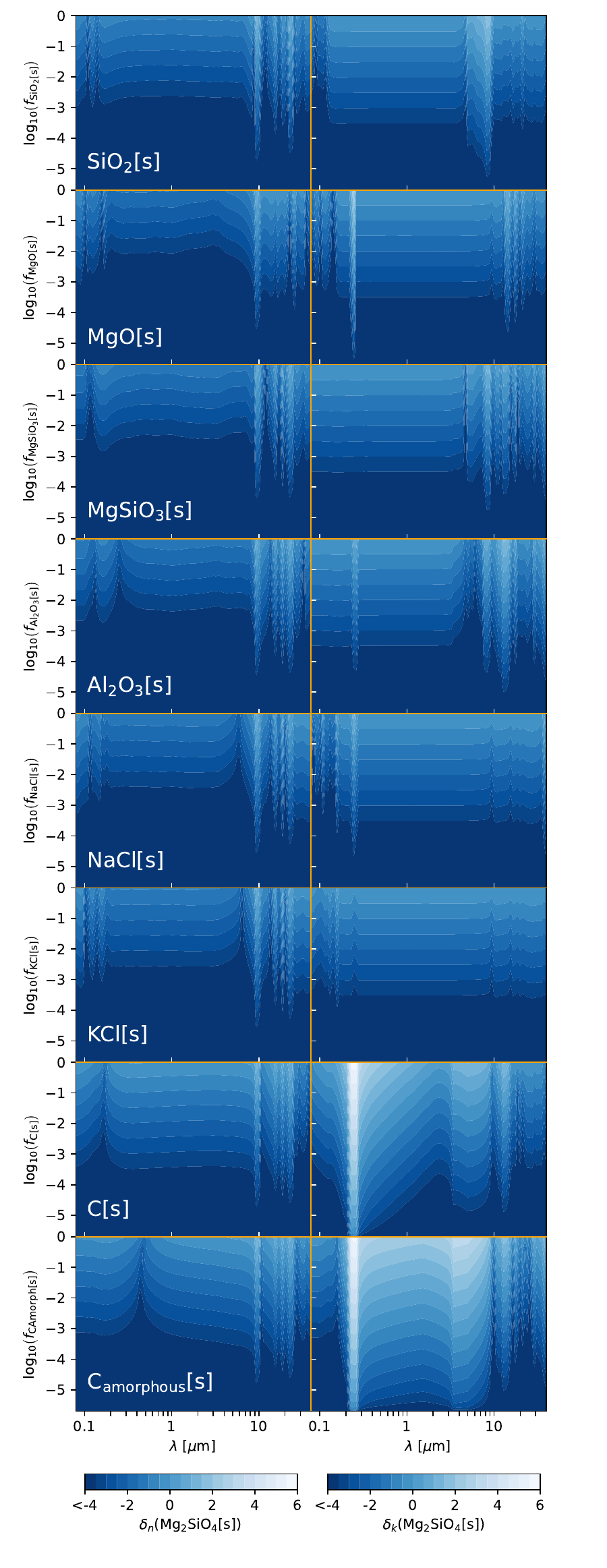}
        \caption{Differences of the real and imaginary part of the refractive index (Eq.~\ref{eq:diff}) for two-component materials compared to the refractive index of Mg$_2$SiO$_4$[s]. The first component is Mg$_2$SiO$_4$[s] and the second component is specified in each plot.}
        \label{fig:mult_mat_hot}
    \end{figure*}

    Clouds in atmospheres of hot Jupiters can be significantly heterogeneous. The most common materials are thought to be iron-, silicon-, or magnesium-bearing species \citep{helling_dust_2006, powell_formation_2018, helling_exoplanet_2023}. In this section, we use two-component materials to analyse the mixing behaviour of different cloud particle materials. While this does not reflect the complexity of cloud particles in a real exoplanet atmosphere, it allows for an analytical study of the wavelength and volume fraction dependence of the effective refractive index. Similar to Sect.~\ref{sec:2comp_emts}, we choose the two-component material \{Mg$_2$SiO$_4$[s], Fe[s]\}. The goal is to determine at which volume fraction Fe[s] inclusions significantly impact the effective refractive index.

    We calculate the logarithm of the absolute relative difference $\delta_n$ of $n_\mathrm{eff}$ compared to $n_\mathrm{Mg_2SiO_4[s]}$ as:
    \begin{align}
        \label{eq:diff}
        \mathrm{\delta_n (\mathrm{Mg_2SiO_4[s]})} = \log_{10} \left( \frac{|n_\mathrm{eff} - n_\mathrm{Mg_2SiO_4[s]}|}{n_\mathrm{Mg_2SiO_4[s]}} \right)
    \end{align}
    and equivalently the absolute relative difference $\delta_k (\mathrm{Mg_2SiO_4[s]})$ of $k_\mathrm{eff}$ compared to $k_\mathrm{Mg_2SiO_4[s]}$. The results for a wavelength range of 0.08~$\mu$m to 39~$\mu$m are shown in Fig.~\ref{fig:fe_mat}. Iron volume fractions of less than 1\% can already increase $k_\mathrm{eff}$ more than 10 times for the wavelength range from $\sim$0.2~$\mu$m to $\sim$10~$\mu$m. The peak difference around $\sim$0.2~$\mu$m is caused by a sudden drop in $k_\mathrm{Mg_2SiO_4[s]}$ values for these wavelengths. This increase can be seen for both LLL and Bruggeman. LLL also shows a large increase for volume fractions below 0.1\% around 4~$\mu$m which is less pronounced for Bruggeman. Overall, these results show that the effective refractive index of cloud particles are strongly affected by the EMT used to calculate it.

    To analyse further two-component materials of cloud forming species in hot Jupiter, we focus on LLL. We consider a bulk material of forsterite with inclusions of various other likely cloud materials. Forsterite ($\mathrm{Mg_2SiO_4[s]}$) is chosen as the bulk material, because it is frequently the dominant cloud particle material in the observable part of the atmosphere \citep{gao_aerosol_2020, helling_exoplanet_2023}. Fig.~\ref{fig:mult_mat_hot} shows the absolute relative differences compared to the refractive index of Mg$_2$SiO$_4$[s]. The same evaluation using Bruggeman instead of LLL can be found in Appendix Fig.~3 of the online material\footnote{Available at https://zenodo.org/records/13373168.}. The results show that all iron-bearing species (Fe[s], FeO[s], Fe$_2$O$_3$[s], Fe$_2$SiO$_4$[s], and FeS[s]) and carbon (C[s] and C$_\mathrm{amorphous}$[s]) strongly impact $k_\mathrm{eff}$ in the wavelength range from approximately $\sim$0.2~$\mu$m to $\sim$10~$\mu$m. TiO$_2$[s] and SiO[s] can also significantly impact $k_\mathrm{eff}$, but mostly around 0.2 $\mu$m which is caused by a sudden drop in $k_\mathrm{Mg_2SiO_4[s]}$ values. The other materials (CaTiO$_3$[s], SiO$_2$[s], MgO[s], MgSiO$_3$[s], Al$_2$O$_3$[s], NaCl[s], and KCl[s]) have a lesser impact on the effective refractive index. While they still can change $k_\mathrm{eff}$ by more than 10 times, they do so only for volume fractions above 10\%. The real part of the effective refractive index $n_\mathrm{eff}$ is mostly affected around 10 $\mu$m. This wavelength corresponds to optical features of Mg$_2$SiO$_4$.

    \subsection{Effective refractive index for clouds in temperate atmospheres}
    \label{sec:2comp_temp}

    \begin{figure}
        \centering
        \includegraphics[width=\hsize]{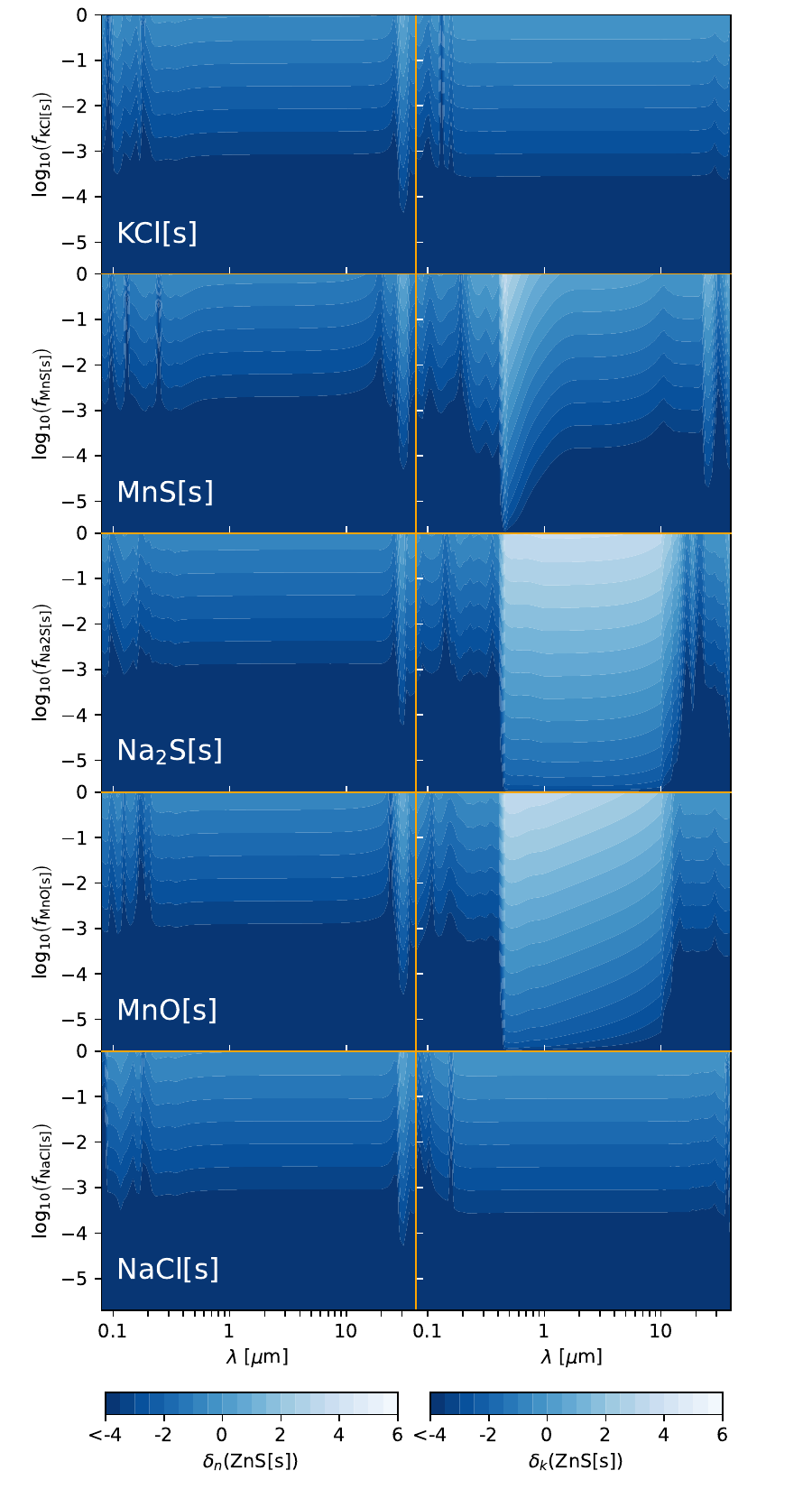}
        \caption{Differences of the real and imaginary part of the refractive index (Eq.~\ref{eq:diff}) for two-component materials compared to the refractive index of ZnS[s]. The first component is ZnS[s] and the second component is specified in each plot.}
        \label{fig:kcl_mat_temp}
    \end{figure}

    In temperate exoplanet atmospheres refractive cloud particle materials, like Mg$_2$SiO$_4$[s], might form cloud layers only below the photosphere of the planet \citep{morley_neglected_2012}. In these environments, cloud particle materials like KCl[s], ZnS[s], and MnS[s] are often discussed as the dominant cloud particle opacity species \citep{mbarek_clouds_2016, gao_microphysics_2018, christie_impact_2022}. We analyse two component materials with ZnS[s] as bulk species and KCl[s], MnS[s], Na$_2$S[s], MnO[s], and NaCl[s] as inclusions. All results are obtained using LLL and expressed in $\delta_n (\mathrm{ZnS[s]})$ and $\delta_n (\mathrm{ZnS[s]})$. Inclusions of KCl[s] and NaCl[s] do not affect the effective refractive index up to volume fractions of 10\%. Sulphur-bearing and MnO[s] inclusions on the other hand can become important at volume fractions of less than 1\%. Overall, our results show that sulphur-bearing species do not have common properties like iron-bearing species. Their impact on heterogeneous cloud optical properties depend on the specific sulphur-bearing cloud particle species.

    \subsection{Non-mixed versus EMT}
    \label{sec:2comp_csvsemt}

    \begin{figure}
        \centering
        \includegraphics[width=\hsize]{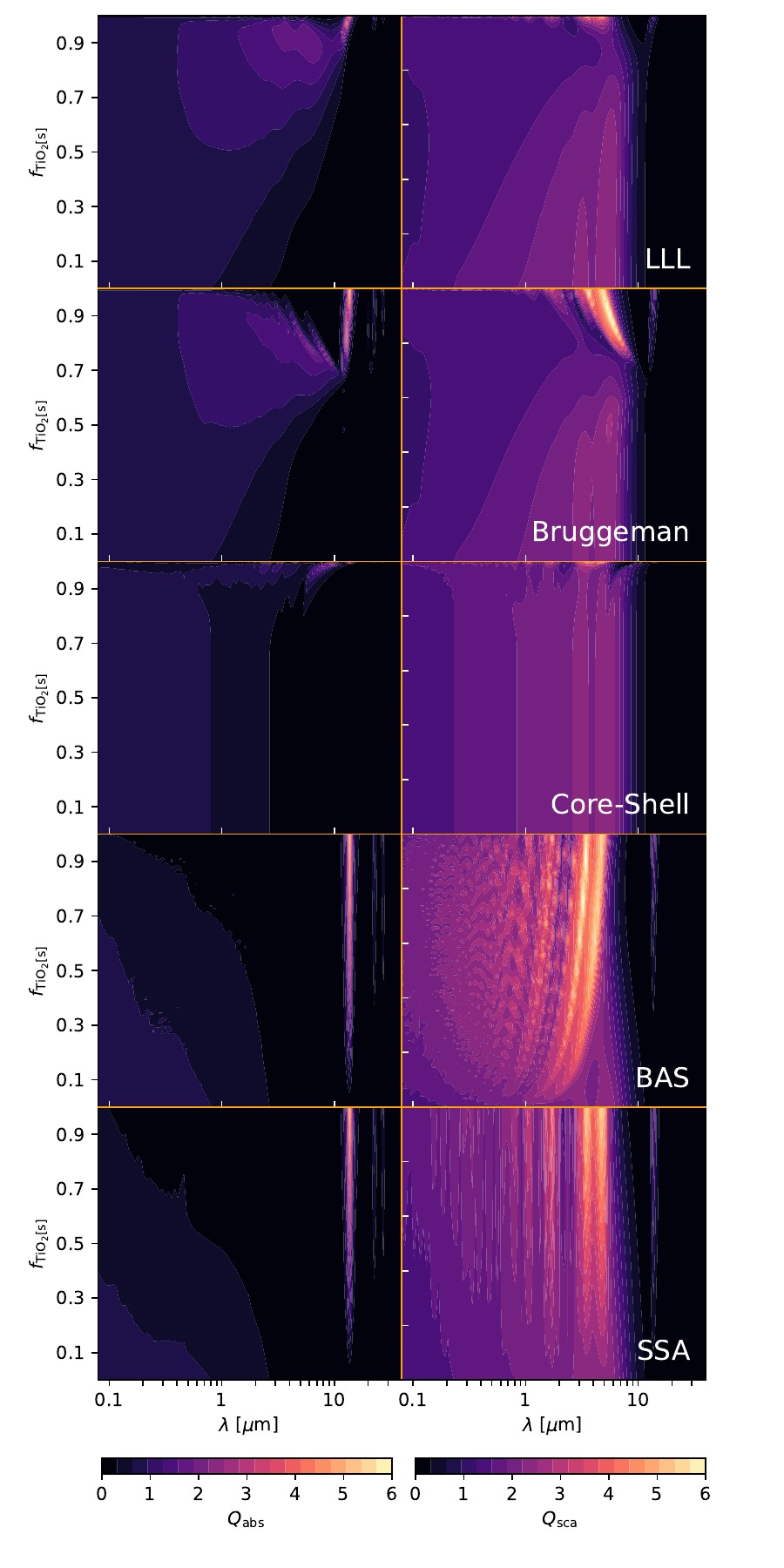}
        \caption{Absorption efficiency $Q_\mathrm{abs}$ (\textbf{Left}) and scattering efficiency $Q_\mathrm{sca}$ (\textbf{Right}) for the two-component material \{TiO$_2$[s],~Mg$_2$SiO$_4$[s]\}.}
        \label{fig:cs_vs_emt}
    \end{figure}

    \begin{figure}
        \centering
        \includegraphics[width=\hsize]{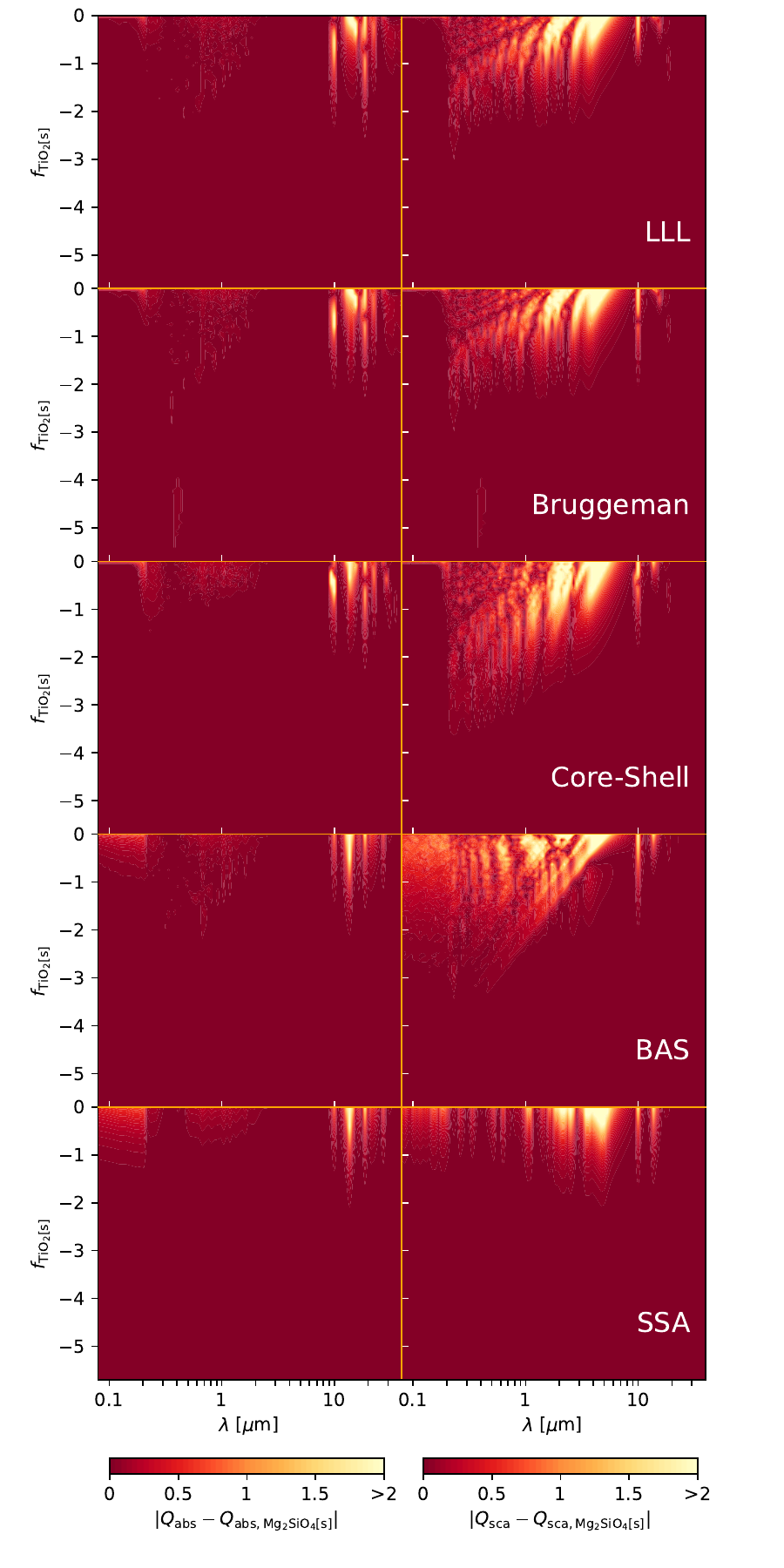}
        \caption{Absolute difference of the absorption efficiency (\textbf{Left}) and scattering efficiency (\textbf{Right}) of the two-component material \{TiO$_2$[s],~Mg$_2$SiO$_4$[s]\} compared to the absorption efficiency $Q_\mathrm{abs, Mg_2SiO_4[s]}$ and scattering efficiency $Q_\mathrm{sca, Mg_2SiO_4[s]}$ of homogeneous Mg$_2$SiO$_4$[s].}
        \label{fig:cs_vs_emt_macro}
    \end{figure}

    In contrast to EMTs, non-mixed treatments assume that cloud particles consist of multiple, homogeneous parts. In this section, we compare the absorption and scattering efficiency for a 1~$\mu$m cloud particle made of TiO$_2$[s] and Fe[s]. TiO$_2$[s] is an often considered as a CCN in hot Jupiters \citep{lee_dust_2015, powell_formation_2018, sindel_revisiting_2022, kiefer_effect_2023, kiefer_fully_2024}. Therefore for the core-shell morphology, we assume that TiO$_2$[s] forms the core and Fe[s] forms the shell. The results for a wavelength range of 0.08~$\mu$m to 39~$\mu$m and TiO$_2$[s] volume fractions between 0 and 1 are shown in Fig.~\ref{fig:cs_vs_emt}.

    The absorption and scattering efficiencies are different for each mixing treatment. At 7~$\mu$m to 8~$\mu$m, the absorption efficiency from LLL, Bruggeman, and core-shell have higher values than either Fe[s] or TiO$_2$[s] in their homogeneous form. This matches the findings from Sect.~\ref{sec:2comp_emts}, where we have shown that mixed cloud particle can have higher absorption efficiencies than homogeneous particles. BAS and SSA on the other hand have their maximum value for homogeneous Fe[s] or TiO$_2$[s]. Furthermore, their absorption efficiency shows a TiO$_2$[s] feature above 10~$\mu$m even at volume fractions below 30\%. The same feature for LLL, Bruggeman, and core-shell vanishes for volume fractions below 70\%. Since heterogeneous particles have inclusions, their effective optical properties are not equivalent to the optical properties of a homogeneous sphere. Therefore, spectral features originating from a spherical geometry might no longer appear in mixed cloud particles. All cloud particles within SSA and BAS on the other hand are perfect homogeneous spheres and therefore can retain the spectral features from the spherical geometry even at low volume fractions.

    For core-shell, the absorption and scattering efficiencies start to be dominated by the Fe[s] even at TiO$_2$[s] volume fractions over 80\%. For all other mixing treatments, TiO$_2$[s] still impacts the absorption and scattering efficiencies even at TiO$_2$[s] volume fractions below 10\%. To analyse if the core can contribute more significantly for other materials, we consider the two component material \{TiO$_2$[s], Mg$_2$SiO$_4$[s]\}. The absolute differences between $Q_\mathrm{abs}$ and $Q_\mathrm{sca}$ compared to the absorption efficiency $Q_\mathrm{abs, Mg_2SiO_4[s]}$ and scattering efficiency $Q_\mathrm{sca, Mg_2SiO_4[s]}$ of homogeneous Mg$_2$SiO$_4$[s] are shown in Fig.~\ref{fig:cs_vs_emt_macro}. For Bruggeman, Core-Shell, and BAS, a TiO$_2$[s] core volume fraction of less than 1\% can already impact the absorption and scattering efficiency by more than 0.5. For SSA on the other hand, the TiO$_2$[s] volume fraction needs to be above 1\% for a change in the the absorption and scattering efficiency of more than 0.5. Similar to Sect.~\ref{sec:2comp}, this means that even spurious materials with a contribution of 1\% or less may have an affect on the opacity of the cloud particles.

\section{Heterogeneous clouds on exoplanets}
\label{sec:planets}

    To explore how choices of EMTs description or particle morphology can effect the interpretation of exoplanet atmosphere observations, the cloud structure of four planets are analysed: HATS-6b \citep{hartman_hats-6b_2015}, WASP-39b \citep{faedi_wasp-39b_2011}, WASP-76b \citep{west_three_2016}, and WASP-107b \citep{anderson_discoveries_2017}.

\begin{figure*}[hbtp!]
        \centering
        \includegraphics[width=0.9\hsize]{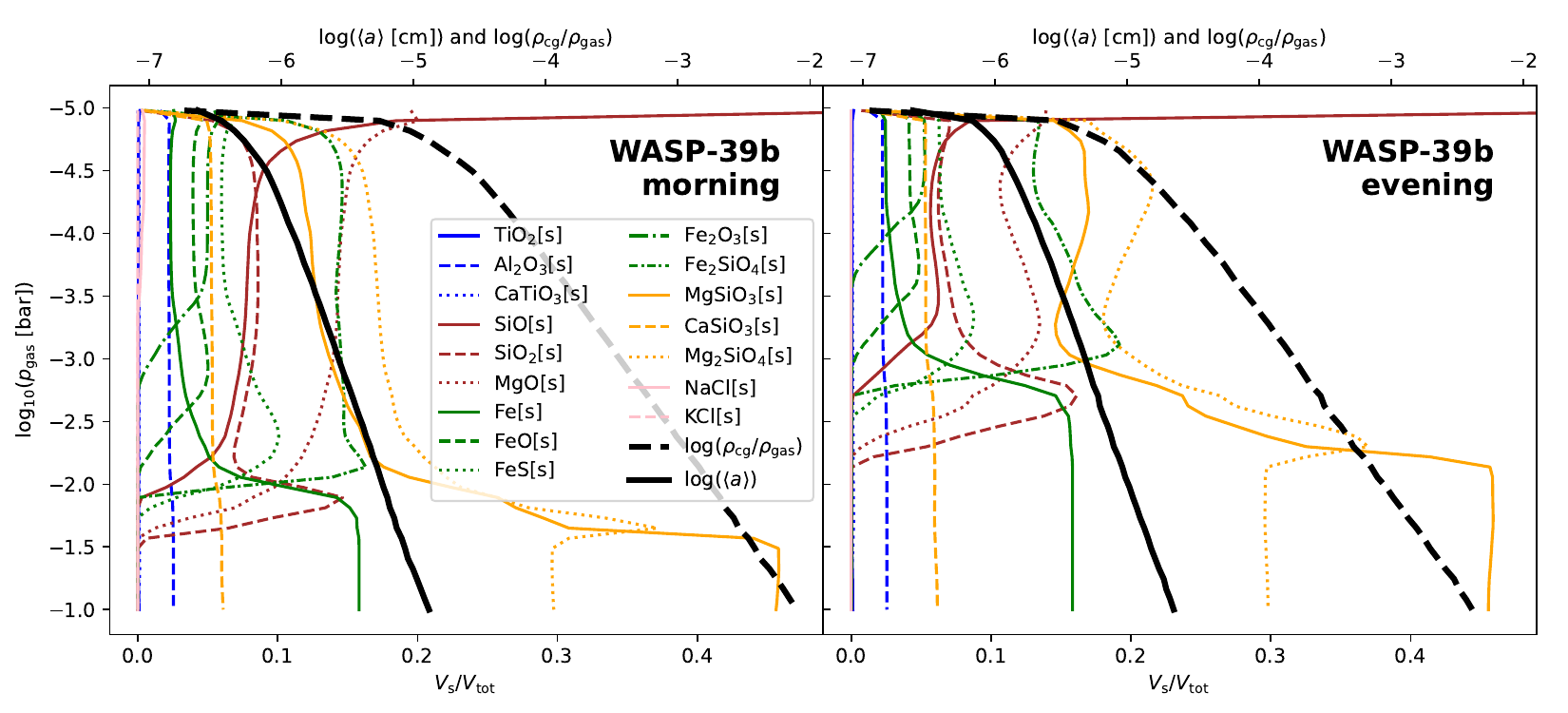}
        \includegraphics[width=0.9\hsize]{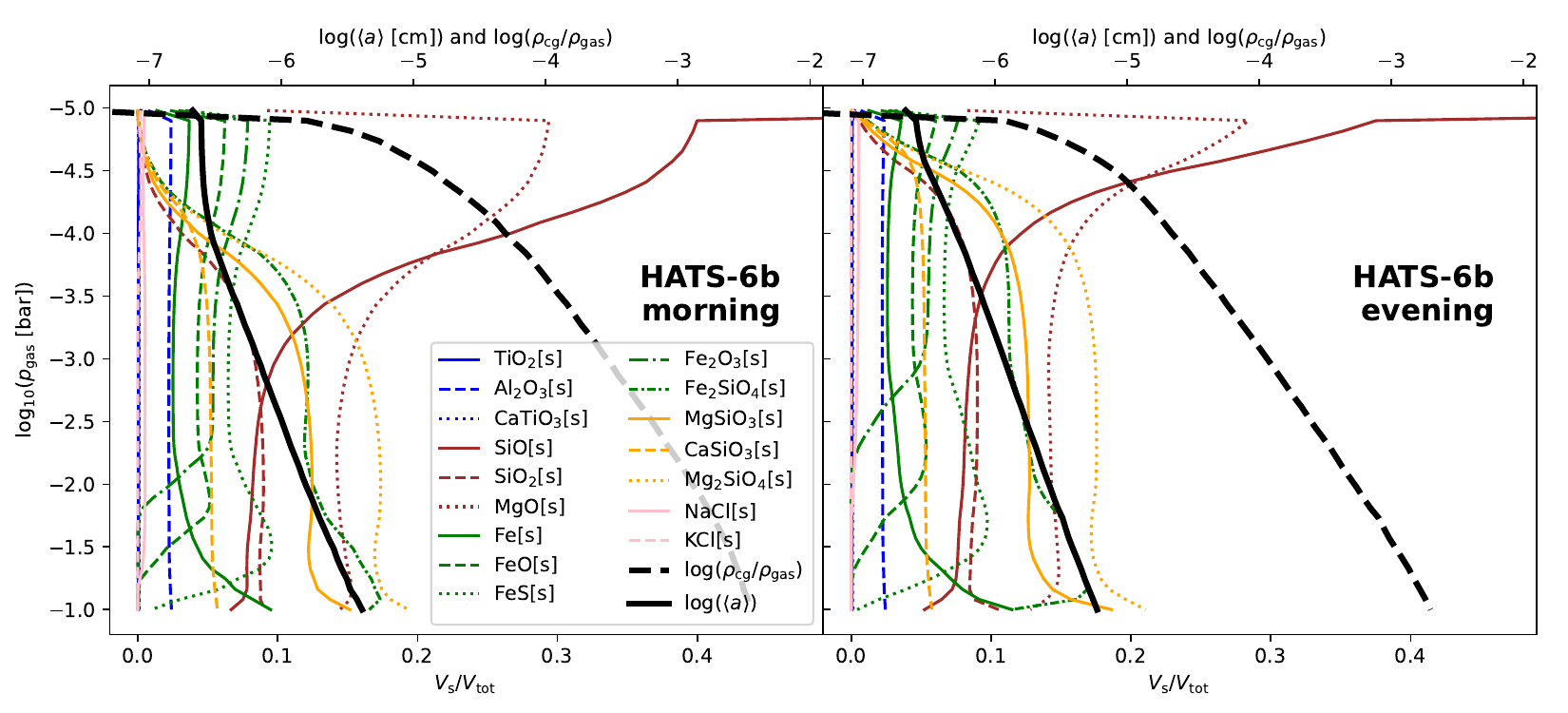}
        \includegraphics[width=0.47\hsize]{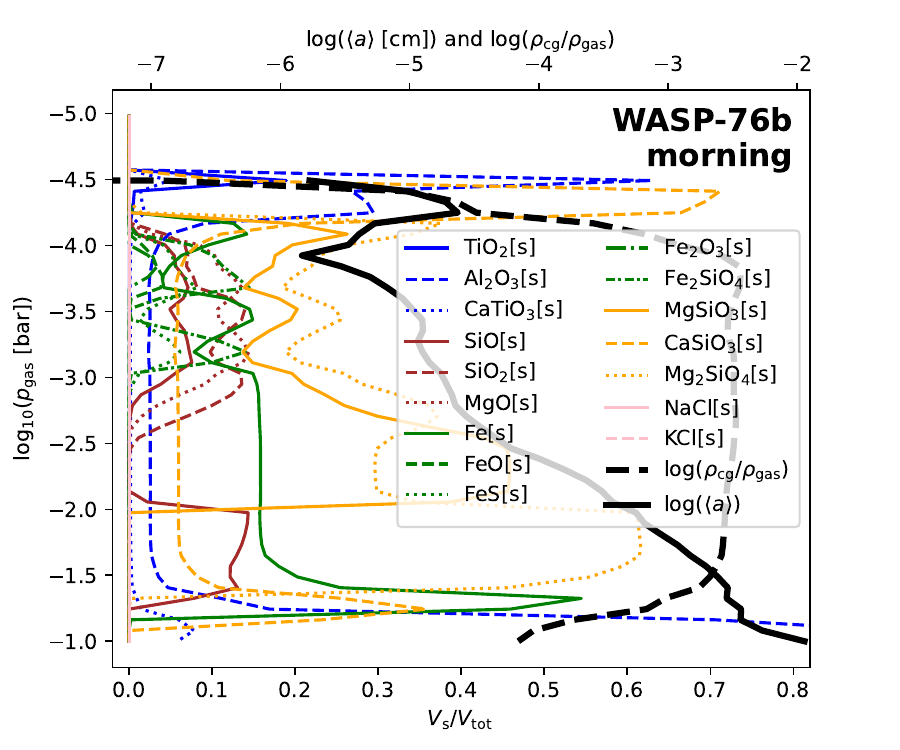}
        \includegraphics[width=0.47\hsize]{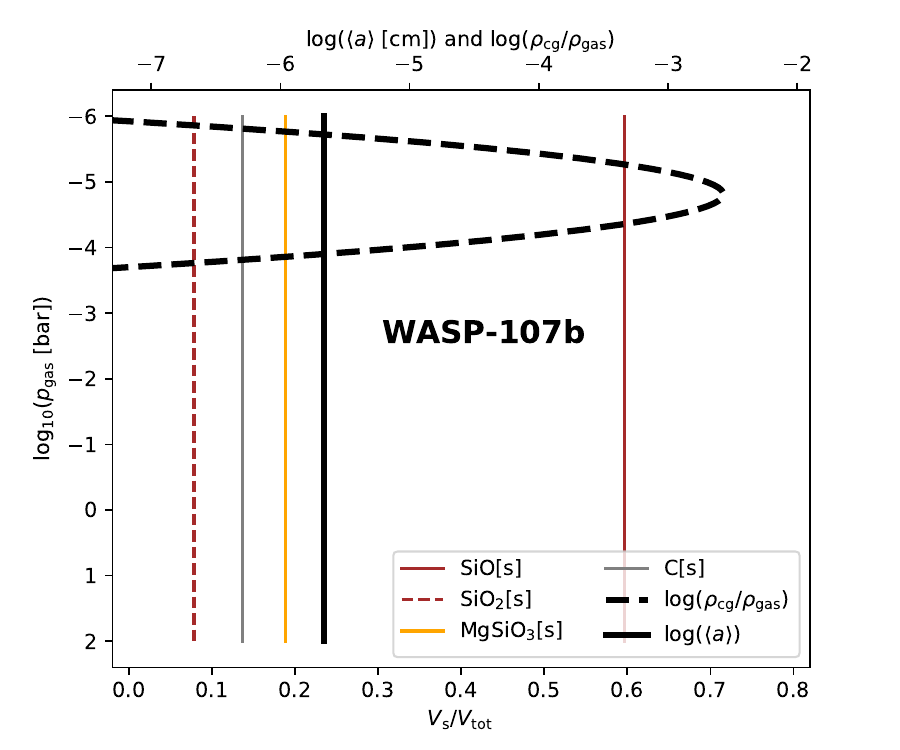}
        \caption{Volume fractions $V_\mathrm{s}/V_\mathrm{tot}$ of each cloud particle species considered (coloured lines), average cloud particle radius $\langle a \rangle$ of all cloud particles, and the total cloud mass fraction $\rho_\mathrm{cloud}/\rho_\mathrm{gas}$ at the equator of the morning and evening terminators. Data taken from \citet{carone_wasp-39b_2023} (WASP-39b), \citet{kiefer_under_2024} (HATS-6b), and \citet{dyrek_so2_2023} (WASP-107b).}
        \label{fig:planet_structs}
    \end{figure*}

    The 1D cloud profiles of WASP-39b, HATS-6b, and WASP-76b at the equator of the morning and evening terminator can be seen in Fig.~\ref{fig:planet_structs}. For WASP-76b, only the morning terminator has clouds and thus the evening terminator is not shown. All cloud profiles are shown in Fig.~6 (HATS-6b), Fig.~7 (WASP-39b), and Fig.~8 (WASP-76b) of the online material\footnote{Available at https://zenodo.org/records/13373168.}. The cloud structure of WASP-107b can be seen in Fig.~\ref{fig:planet_structs} as well. This cloud structure was derived through a one dimensional ARCiS retrieval rather than forward modelling and thus considers fewer cloud particle materials and assumes constant volume fractions with pressure.

    \subsection{Pressure dependent cloud optical properties of WASP-39b}
    \label{sec:planets_hats6b}

    \begin{figure}
        \centering
        \includegraphics[width=\hsize]{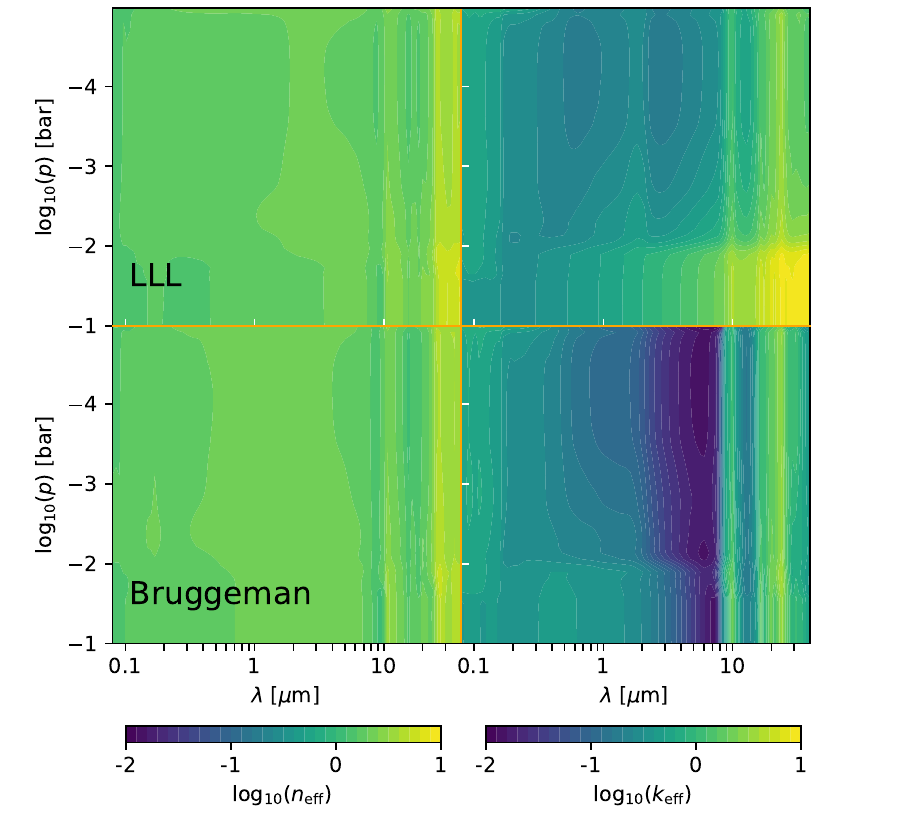}
        \caption{Effective refractive index at the equator of the morning terminator of WASP-39b.}
        \label{fig:planet_emt}
    \end{figure}
    \begin{figure}
        \centering
        \includegraphics[width=\hsize]{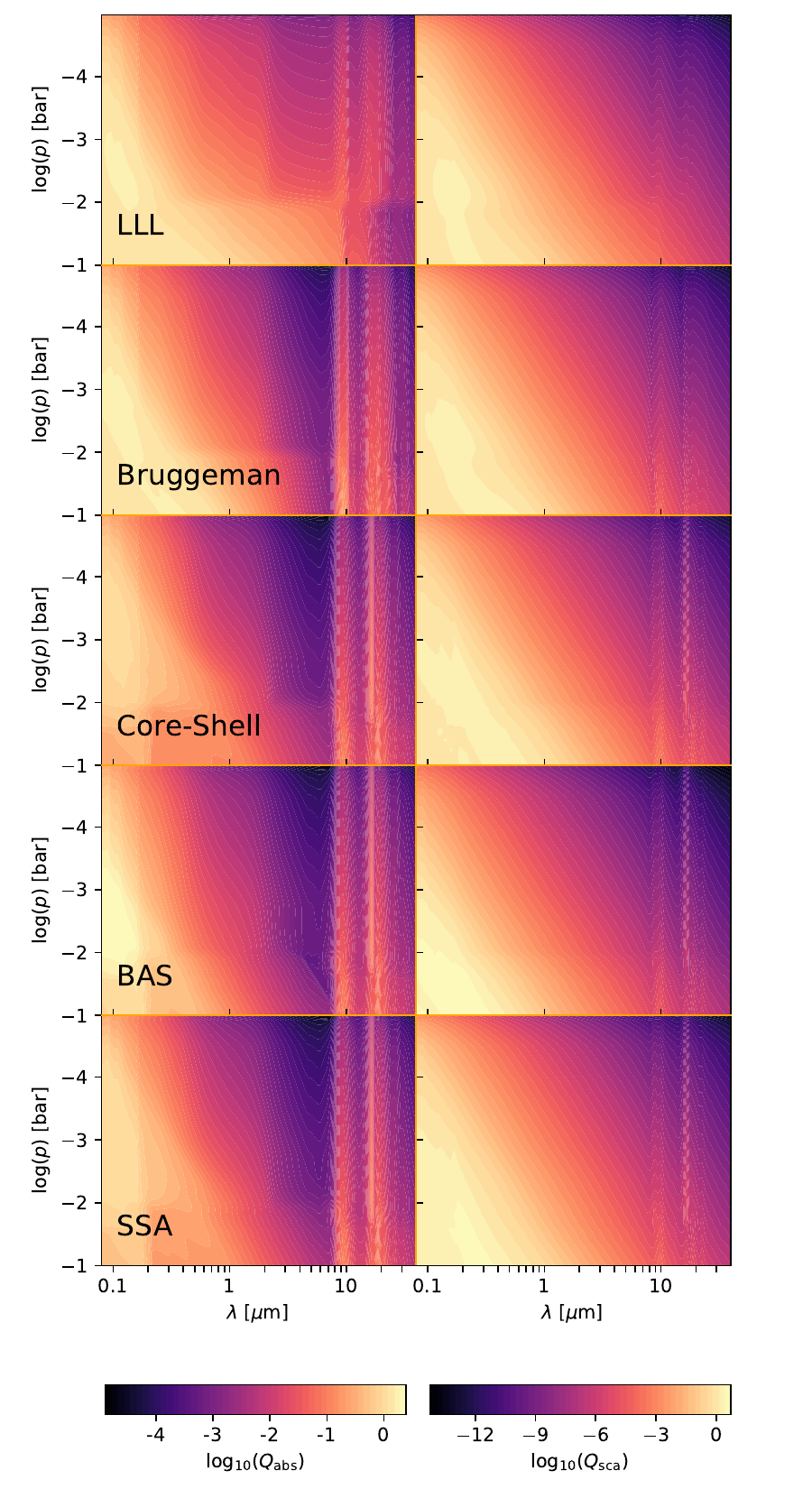}
        \caption{Absorption (\textbf{Left}) and scattering efficiency (\textbf{Right}) of WASP-39b at the equator of the morning terminator.}
        \label{fig:planet_mie}
    \end{figure}

    To investigate the pressure dependent refractive index, absorption efficiency and scattering efficiency, the cloud structure of WASP-39b at the equator of the morning terminator was selected\footnote{The figures of HATS-6b, and WASP-76b can be seen in Fig.~4 and 5 of the online material which is available at https://zenodo.org/records/13373168.}.

    At the top of the atmosphere around 10$^{-5}$~bar, the cloud particle composition is dominated by SiO from nucleation. Going deeper into the atmosphere, the volume fraction of other species rapidly increases due to the increasing density. Cloud particles remain highly mixed throughout the atmosphere with Mg$_2$SiO$_4$[s], MgSiO$_3$[s], MgO[s] and Fe$_2$SiO$_4$ being the dominant materials. Other materials are present at lower volume fractions. There is a rapid change in material composition around 10$^{-2}$~bar where the magnesium-silicates enstatite (MgSiO$_3$) and forsterite (Mg$_2$SiO$_4$) become more dominant. This change happens at the same pressure where Fe$_2$SiO$_4$[s] evaporates and Fe[s] becomes the dominant iron-beraing cloud particle material. Presumably the liberation of oxygen from Fe$_2$SiO$_4$ leads to the Mg:O 1:3  stoichiometry (enstatite)  to become favoured over the Mg:O 1:2 stoichiometry (forsterite). This effect is further amplified by the evaporation of SiO$_2$ around the same pressure layer which liberates additional oxygen. Throughout the atmosphere, the average particle size and cloud particle mass load increases steadily with increasing pressure due to bulk growth. The same is broadly true for the evening terminator, with the major difference that the evaporation of Fe$_2$SiO$_4$[s] happens at a lower pressure (higher altitude), closer to 10$^{-3}$~bar. This is due to higher temperatures of the evening terminator in general. The effective refractive index from different EMTs throughout the morning terminator is shown in Fig.~\ref{fig:planet_emt}.

    All EMTs show the highest $n_\mathrm{eff}$ and $k_\mathrm{eff}$ values for pressures higher than $10^{-2}$~bar. This coincides with the change from Fe$_2$SiO$_3$[s] to Fe[s] being the dominant iron-bearing species. As we have seen in Sect.~\ref{sec:2comp_hot}, Fe[s] has a stronger effect on the effective refractive index than Fe$_2$SiO$_3$[s]. This increase is much stronger for LLL than for Bruggeman. For pressures lower than 10$^{-2}$~bar (higher altitudes) a slight increase in refractive index values can be seen which is due to SiO[s] being the dominant material. While this is a consequence of the upper boundary condition of the model, the changes in refractive index are small and unlikely to be significant in transmission spectra calculations.

    Between $10^{-4}$~bar to 10$^{-2}$~bar, the refractive index values are generally lower. Compared to LLL, Bruggeman predicts lower values for $k_\mathrm{eff}$, especially between 3~$\mu$m to 8~$\mu$m. As is shown in Sect.~\ref{sec:planets_trans}, this results in a "window" where cloud particles are less opaque. This is a direct consequence of the reduced impact of inclusions made from iron-bearing species within Bruggeman compared to LLL and shows the impact of the choice of EMT.

    To further analyse the impact of mixing treatments on cloud optical properties, the absorption and scattering efficiencies for Bruggeman and LLL as well as the non-mixed treatments core-shell, BAS, and SSA are shown in Fig.~\ref{fig:planet_mie}. The absorption efficiencies of Bruggeman, core-shell particles, BAS and SSA are similar for all pressures and wavelengths up to 8~$\mu$m. Only LLL shows clear deviations around 1~$\mu$m to 8~$\mu$m where $Q_\mathrm{abs}$ is larger compared to the others. The non-mixed treatments show silicate features around and above 10~$\mu$m. The same features can be seen in LLL and Bruggeman but are weaker in comparison. This indicates that non-mixed treatments might retain more spectral features from the individual cloud particle materials than EMTs.

    \subsection{Transmission Spectrum}
    \label{sec:planets_trans}

    \begin{figure*}[hbtp!]
        \centering
        \includegraphics[width=\hsize]{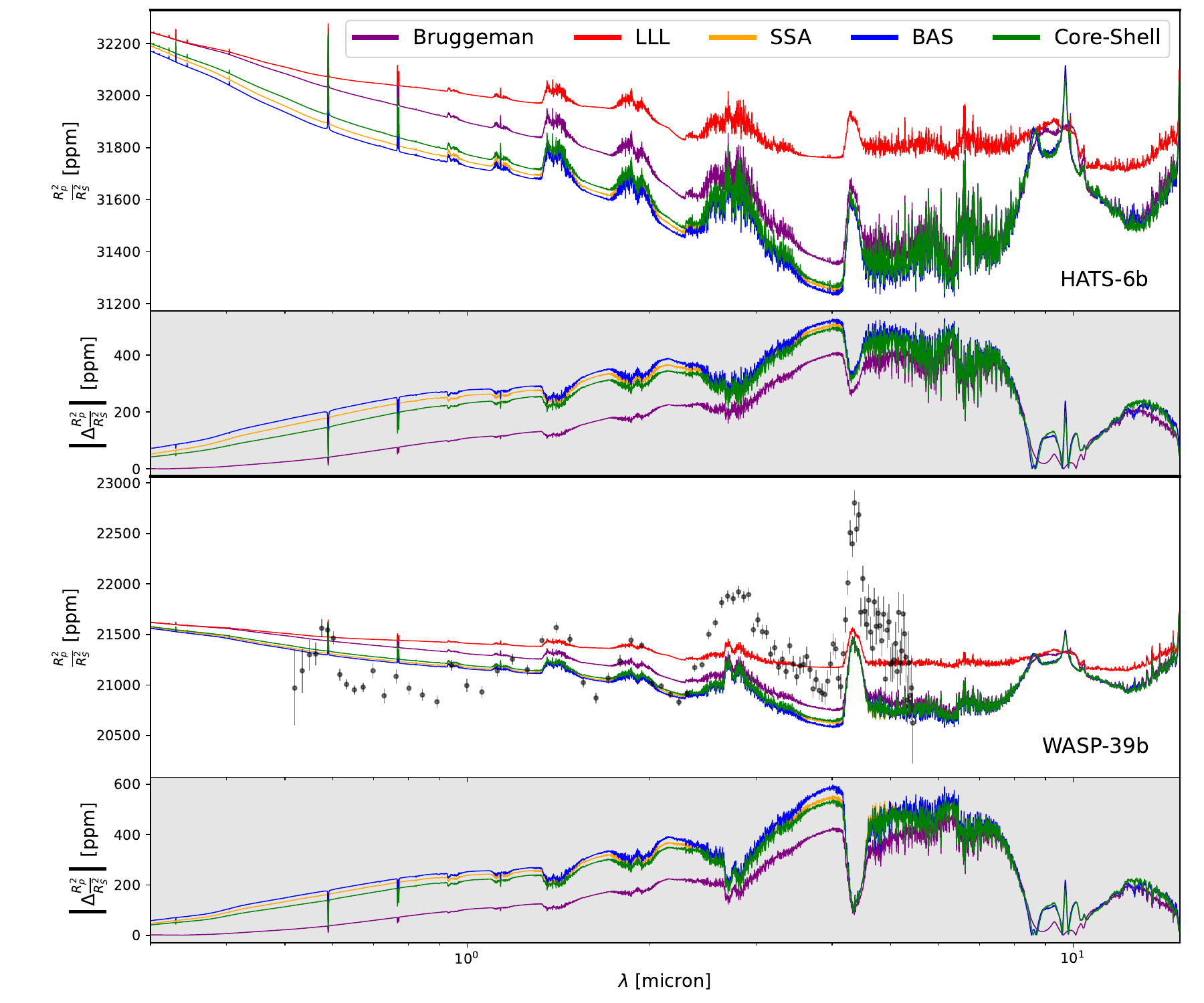}
        \caption{Transmission spectra of different planets using EMTs and non-mixed treatments. The absolute difference to LLL are shown underneath each plot. In both figures, Core-Shell, BAS, and SSA are overlapping. For WASP-39b, JWST NIRspec Prism data from \citet{rustamkulov_early_2023} is shown.}
        \label{fig:p_spectra}
    \end{figure*}

    To investigate the impacts of EMTs and non-mixed treatments on transmission spectra, we choose a wavelength range of 0.3 $\mu$m to 15~$\mu$m which covers the JWST instruments NIRspec Prism and MIRI LRS. In this section, we show the transmission spectrum of HATS-6b and WASP-39b calculated using different mixing treatments (Fig.~\ref{fig:p_spectra}). The results of WASP-39b are compared to the observations of \citet{rustamkulov_early_2023}. An offset of -867~ppm was applied to all synthetic spectra of WASP-39b to achieve the best fit of LLL with the data. The planets WASP-76b and WASP-107b are analysed in more detail in Sect.~\ref{sec:planets_wasp76b} and \ref{sec:planets_wasp107b}, respectively. To analyse the differences between EMTs, the absolute differences of the transit depth compared to LLL are calculated:
    \begin{align}
        \left| \Delta \frac{R^2_p(\lambda)}{R^2_s (\lambda)} \right| = \left| \frac{R^2_p(\lambda) - R^2_\mathrm{p,LLL}(\lambda)}{R^2_s(\lambda)} \right|
    \end{align}

    Both HATS-6b and WASP-39b have cloud particle radii of less than 0.1~$\mu$m between 10$^{-5}$~bar to $10^{-2}$~bar. Their cloud mass fraction steadily increases with pressure and reaches more than $10^{-4}$ at pressures higher than $10^{-4}$~bar. Hence, the transmission spectra of both planets is affected by the cloud particle opacities. For WASP-39b and HATS-6b the differences are up to 500~ppm.

    The three non-mixed treatments Core-Shell, BAS and SSA produce nearly identical spectra for both planets. The main difference in the opacity calculation between SSA and BAS is that SSA divides the cloud particle materials into larger, but fewer particles and BAS into smaller, but more particles. This is in particular interesting since changes in size can impact the spectral features of cloud particles \citep{wakeford_transmission_2015}. The number density of cloud particles, however, only impacts the overall amount of light that gets absorbed but not the shape of spectral features. While these differences could impact the transmission spectrum, we see little differences between BAS and SSA in both exoplanets. Core-Shell differs from BAS and SSA by considering interactions between the core and the shell. As we showed in Sect.~\ref{sec:2comp_csvsemt}, this produces features in the absorption and scattering efficiency which could result in spectral features in the transmission spectrum. However, no such features are seen in both planets because the cloud particle composition is dominated by the growth material which forms the shell, not the nucleation that forms the core. This matches the results from \citet{powell_transit_2019} who found contributions of TiO$_2$[s] cores within Fe[s], Mg$_2$SiO$_4$[s], and Al$_2$O$_3$[s] shells negligible.

    For both planets, non-mixed cloud particles exhibit stronger spectral features around 8~$\mu$m to 10~$\mu$m than well-mixed cloud particles. In particular a large feature from Mg$_2$SiO$_4$[s] can be seen in core-shell, BAS, and SSA but not in Bruggeman,  or LLL. This matches the findings from Sect.~\ref{sec:planets_hats6b} where non-mixed approximations showed larger silicate features in the absorption and scattering efficiencies.

    In both exoplanets, Bruggeman, Core-Shell, SSA, and BAS predict a lower transmission depth than LLL. This can be explained by the high refractive index of iron-bearing species which strongly impacts LLL (see Sect.~\ref{sec:2comp_hot}). Even when silicate species are the dominant cloud material, the large values for the imaginary part of the refractive index $k$ of iron-bearing species dominates the effective refractive index calculation. Within SSA and BAS, iron-bearing species are assumed to form their own homogeneous cloud particles. As we have shown in Sect.~\ref{sec:2comp_emts}, homogeneous iron-bearing species can have lower absorption efficiency than mixed particles. Similarly within Core-Shell, the iron bearing species form a homogeneous shell and thus do not increase the absorption efficiency as much.

    All 5 methods produce a featureless spectrum at wavelengths below 1~$\mu$m in HATS-6b and WASP-39b. This effect is most clearly shown in the narrow sodium and potassium lines. For WASP-39b, we can compare our results to the observations of \citet{rustamkulov_early_2023}. We find that all our spectra show less molecular features than the observations. This indicates that WASP-39b might have less clouds than predicted by the model. LLL shows larger differences to the observations than the other mixing treatments. However, WASP-39b shows that it is difficult to disentangle the effect of  cloud particle mixing treatments and the amount of cloud coverage. Observations of cloud spectral features might help to gain further insights. In our results, the non-mixed treatments Core-Shell, BAS, and SSA produce sharper cloud particle features than LLL and Bruggeman. Cloud particle features are most relevant around 10~$\mu$m. This is in agreement with previous theoretical studies \citet{wakeford_transmission_2015} and with the detections of silicate features by \citet{grant_jwst-tst_2023} and \citet{dyrek_so2_2023}.

    \subsection{Iron-bearing cloud particle materials in WASP-76b}
    \label{sec:planets_wasp76b}

    \begin{figure}
        \includegraphics[width=\hsize]{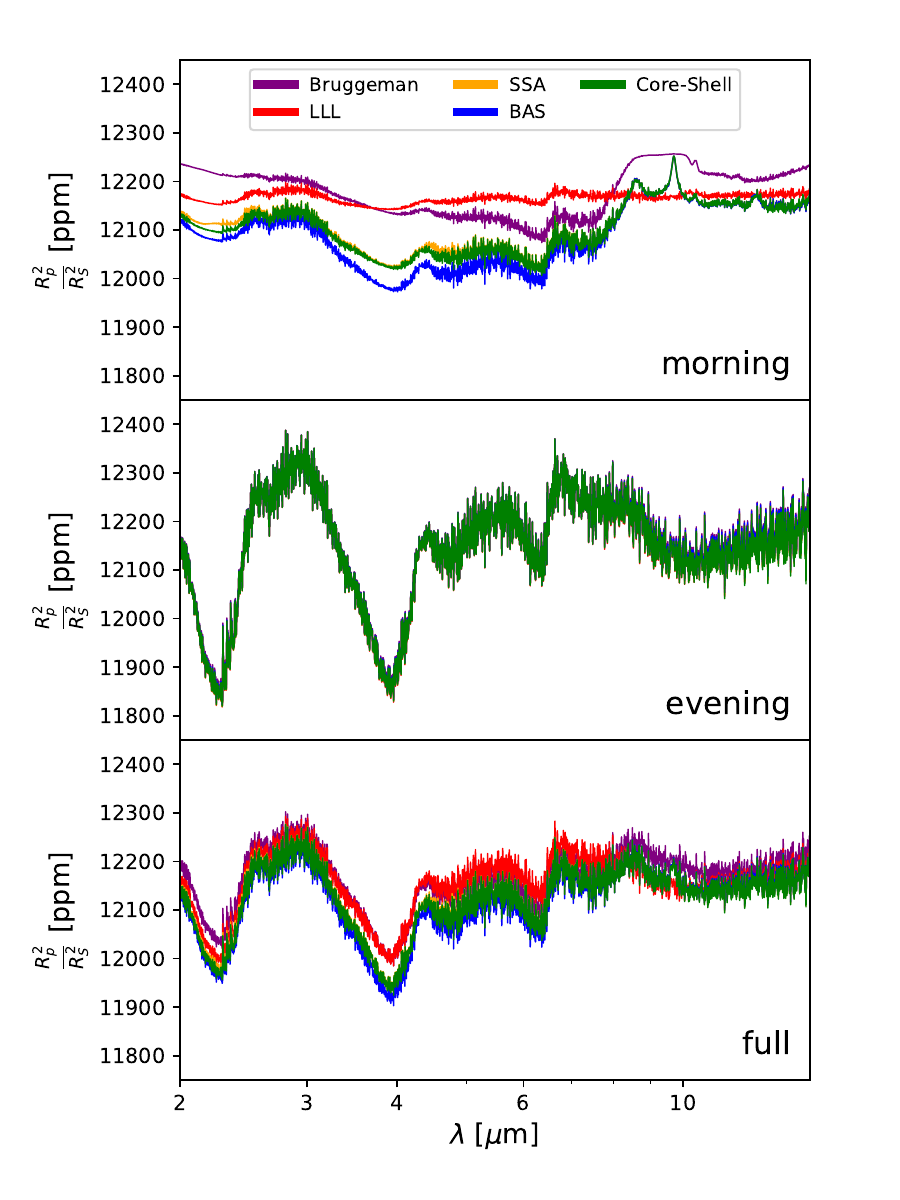}
        \caption{Transmission spectra of WASP-76b by considering the contributions of all terminator profiles (full), only the morning terminator profiles (morning), and only the evening terminator profiles (evening). Because the evening and morning terminator only cover half of the terminator region each, there transit depths were multiplied by 2 to be comparable to the full transit spectrum.}
        \label{fig:wasp76b}
    \end{figure}

    WASP-76b is an ultra hot Jupiter \citep{west_three_2016} which is a type of exoplanet that is typically predicted to have cloud free day-sides and cloudy night-sides \citep{helling_exoplanet_2023, demangeon_asymmetry_2024}. Several high resolution observations have confirmed the presence of gas-phase iron, vanadium, magnesium, and sodium species \citep{ehrenreich_nightside_2020, pelletier_vanadium_2023, gandhi_retrieval_2023, maguire_high_2024}. The observations of \citet{ehrenreich_nightside_2020} found neutral iron in the morning limb but not in the evening limb. This detection was followed up by \citet{savel_no_2022} using post processed GCMs. They found that the differences between the limbs can be explained by the impact of clouds on the observations. WASP-76b is thus an interesting target to study cloud particle morphologies since it has iron in the atmosphere and is expected to have a cloudy morning terminator. This is in agreement with our modelling results. We find no clouds at the equator of the evening terminator and only few cloud particles at higher latitudes. The morning terminator has cloud coverage for all latitudes and their composition includes iron-bearing species. To derive the cloud structure, \citet{savel_no_2022} used an equilibrium thermal stability criteria. As a result, they do not predict iron clouds in the uppermost atmosphere because iron is already thermally stable in deeper layers. In contrast, our cloud model is kinetic which allows to consider the micro-physics of cloud formation and therefore it is possible for cloud particle materials to grow in regions outside where an equilibrium model would predict.

    The full transmission spectrum of WASP-76b is produced by considering the contribution of all latitudes along both terminators. For the morning and evening terminator spectrum, only the contributions of lat = -90$^\circ$ and lat = 90$^\circ$ are considered, respectively. Since each limb only covers half the planet, the transit depth values of morning and evening only spectra were doubled to be comparable to the full transit spectrum. The transit depth of the morning limb, evening limb, and full transmission spectrum can be seen in Fig.~\ref{fig:wasp76b}. The evening terminator spectrum shows no cloud features and no significant muting of the molecular lines. Both EMTs and non-mixed treatments predict nearly an identical transmission spectra despite the presence of some high latitude clouds. The morning terminator spectrum shows clear signs of clouds. In agreement with Sect.~\ref{sec:planets_trans}, LLL leads to the flattest spectra. BAS leads to the least flat spectrum. The difference in transit depth between LLL and BAS is maximum $\sim$200~ppm.  Bruggeman, core-shell, and SSA are between the other methods. Similar to Sect.~\ref{sec:planets_trans}, SSA and core-shell have a nearly identical transmission spectrum. All techniques other than SSA and core-shell are separated by at least $\sim$50~ppm to each other. Around 10~$\mu$m, the non-mixed treatments show clear Mg$_2$SiO$_4$[s] spectral features. Bruggeman shows a broad feature and LLL predicts only a flat spectrum. The full transit spectrum still exhibits differences between well-mixed and non-mixed cloud particles. However, the differences in the flatness of the spectrum is much less pronounced than in the morning limb alone. The spectral features of Mg$_2$SiO$_4$[s] can no longer be seen for any method for the full transit spectrum.

    Our results show that the EMTs and three non-mixed treatments show clear differences in the flatness of the spectrum and the cloud particle spectral features around 10~$\mu$m. Detailed limb asymmetry observations of WASP-76b thus might allow the investigation of cloud particle morphologies at the morning terminator, removing the dilution of the effect by a mostly cloudless evening terminator.

    \subsection{Cloud detections in WASP-107b}
    \label{sec:planets_wasp107b}

    \begin{figure}
        \includegraphics[width=\hsize]{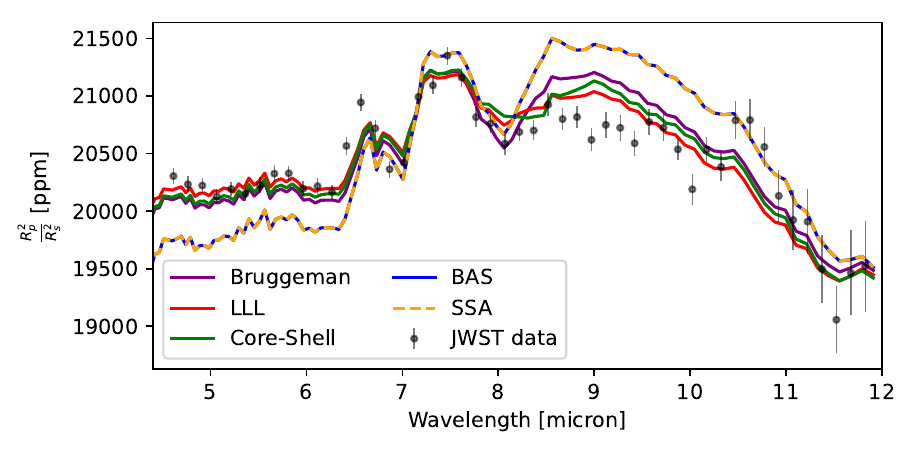}
        \caption{Transmission spectra of WASP-107b reproduced from the cloud structure retrieved by ARCiS. Cloud structure and JWST MIRI data taken from \citet{dyrek_so2_2023}.}
        \label{fig:wasp107b}
    \end{figure}

    \citet{dyrek_so2_2023} detected silicon-bearing clouds in the atmosphere of WASP-107b. Within their ARCiS retrieval \citep{min_arcis_2020}, they assumed heterogeneous cloud particles made from SiO[s], SiO$_2$[s], MgSiO$_3$[s], and C$_\mathrm{amorphous}$[s]. They assumed well-mixed particles and calculated the effective refractive index using Bruggeman. Furthermore, they considered a distribution of hollow-spheres \citep[DHS;][]{min_modeling_2005}. It is important to note that the cloud modelling used for WASP-107b is therefore different than the modelling used for HATS-6b, WASP-39b, and WASP-76b.

    In this section, we reproduce the results of the ARCiS retrieval using the best fit parameters from \citet{dyrek_so2_2023}. This fit includes the temperature-pressure profile, gas-phase abundances, and cloud particle properties. The cloud structure is shown in Fig.~\ref{fig:planet_structs}. For the gas-phase, we consider the following opacity species:
    H$_2$O, CO, CO$_2$,
    CH$_4$ \citep{yurchenko_hybrid_2017},
    SO$_2$ \citep{underwood_exomol_2016},
    H$_2$S \citep{azzam_exomol_2016},
    NH$_3$ \citep{coles_exomol_2019},
    SiO \citep{barton_exomol_2013},
    PH$_3$ \citep{sousa-silva_exomol_2015},
    HCN \citep{barber_exomol_2014},
    and C$_2$H$_2$ \citep{chubb_exomol_2020}.
    We compare the differences in the transit spectrum caused by EMTs (Bruggeman and LLL) and non-mixed treatments (core-shell, BAS, and SSA). For core-shell, SiO[s] was assumed as the core material. In contrast to \citet{dyrek_so2_2023}, we do not consider a DHS. The results are shown in Fig.~\ref{fig:wasp107b}. Because the cloud particle properties were retrieved using Bruggeman, the reference pressure and pressure range of the transmission spectrum calculation were chosen to minimise the differences to the observational data when using Bruggeman. However, our solution still slightly differs to the ARCiS fit from \citet{dyrek_so2_2023}. This can be explained by the fact that ARCiS is a retrieval model that searchers for the best fit parameters for its set-up. Differences in our set-up for transmission spectrum calculations compared to theirs therefore result in slightly different transmission spectra. For all other mixing treatments, the same reference pressure and pressure range as for Bruggeman was used. Afterwards, an offset was applied to the transmission spectra from each mixing treatment except Bruggeman to achieve the best fit to the data.

    Between the different mixing treatment there are transit depth differences of up to 500~ppm around 9~$\mu$m. The differences are mainly caused by C$_\mathrm{amorphous}$[s]. As we have shown in Sect.~\ref{sec:2comp_hot}, C$_\mathrm{amorphous}$[s] has a similarly effect on cloud particle optical properties than iron-bearing species. Similar to Sect.~\ref{sec:planets_trans} and \ref{sec:planets_wasp76b}, LLL leads to the overall flattest transmission spectrum. SSA and BAS have the largest differences compared to the other mixing treatments while for all other planets both were close to core-shell. This difference can be explained by a large volume fraction of the core species ($f_\mathrm{SiO} = 0.906$). For the other planets, such high SiO[s] concentrations were only present at the top of the atmosphere where they impact the transmission spectrum less.

    Overall, our results show that the differences in the optical properties between the mixing treatments have an observable effect. Depending on which treatment is used, different solutions to the atmospheric structure might be reached (See Sect~\ref{sec:dis_howyoushould}).

\section{Discussion}
\label{sec:dis}

    Our results show that the choice of EMT or non-mixed theory can have an observable impact on transmission spectra of cloudy atmospheres with heterogeneous cloud particles. The differences between mixing treatments is discussed in Sect.~\ref{sec:dis_compare}. The impact of mixing treatments on transmission spectra is analysed in Sect.~\ref{sec:dis_howyoushould}. Lastly, in Sect.~\ref{sec:dis_iron}, the importance of carbon and iron-bearing species as cloud particle material are discussed.

    \subsection{Comparing mixing treatments}
    \label{sec:dis_compare}

    For non-mixed particles, our results show little difference between core-shell, BAS and SSA in most transmission spectra produced for this study. Only at high volume fractions of the core species (e.g. for WASP-107b) does SSA and BAS produce different transit spectra than core-shell. The similar results in the other cases can be explained with how the cloud particle materials are separated depending on the mixing treatment. Dominant species, like Mg$_2$SiO$_4$[s], have large enough volume fractions that neither the size change of BAS, the number density change of SSA, nor a small core made from a different material significantly impacts their contribution to the optical property of the cloud layer. Non-dominant species on the other hand, like MgO, have low number densities in SSA, small sizes in BAS, and small contributions to the core-shell calculations. This makes their contributions to the cloud particle optical properties negligible.

    If cloud particles are assumed to be well-mixed, EMTs have to be used to describe their optical properties. The most common EMTs used to study mixed materials in exoplanets are Bruggeman and LLL (see Table~\ref{tab:summary}). However, our analysis of two-component materials (see Sect.~\ref{sec:2comp_emts}) and transit spectra (see Sect.~\ref{sec:planets}) show that the optical properties of cloud particles can differ significantly between Bruggeman and LLL. This difference is mainly caused by species with a high imaginary part of the refractive index, like iron-bearing species or carbon. Laboratory experiments of mixed materials have shown that either Bruggeman or LLL can be more accurate, depending on the material and wavelength \citep[see e.g.][]{kolokolova_scattering_2001, voshchinnikov_effective_2007, thomas_investigations_2009}. For clouds in exoplanet atmospheres, it is unknown if Bruggeman or LLL is more accurate. A big advantage of LLL over Bruggeman is the computation time. Bruggeman requires a computationally intensive minimisation algorithm whereas LLL presents an analytical solution which is quick to execute. Many larger frameworks, like GCM, thus either use LLL \citep{lee_modelling_2023}, or a combination of LLL and Bruggeman \citep{lee_dynamic_2016, lines_exonephology_2018}. Smaller frameworks, like transmission spectrum calculations, can afford the slower computational times and can use Bruggeman \citep{min_arcis_2020, dyrek_so2_2023}. While the choice between LLL or Bruggeman, motivated by computational time makes sense, it also has implications on the optical properties of the clouds and thus on the derived temperature structure from observations, the calculation of opacities in forward models, and on transmission spectra of the exoplanet.

    \subsection{How do assumptions on heterogeneous cloud particles affect predicted observables?}
    \label{sec:dis_howyoushould}

    The optical properties of heterogeneous cloud particles depends on their shape, composition, and material distribution. However, in exoplanet atmospheres, only few cloud particle properties can currently be precisely determined from observations \citep[e.g.][]{grant_jwst-tst_2023, dyrek_so2_2023}. Therefore multiple assumptions have to be made to calculate cloud particle optical properties. The validity of these assumptions is hard to prove and most often depends on the underlying physics of cloud formation.

    Many observations of flat transmission spectra in the optical wavelength range have been explained by clouds \citep[e.g.][]{bean_ground-based_2010, kreidberg_clouds_2014, espinoza_access_2019, spyratos_transmission_2021, libby-roberts_featureless_2022}. Our results for HATS-6b and WASP-39b show that all mixing treatments predict strongly muted spectra in the optical. Therefore, observations in optical wavelengths on their own are not suited to study cloud particle morphologies. Observing the spectral features of cloud particle materials in the infrared allows more detailed insights. \citet{grant_jwst-tst_2023} found in their observations of WASP-17b a peak between 8~$\mu$m to 9~$\mu$m which was linked to SiO$_2$[s] clouds. Furthermore, they argue that this spectral feature is better explained with crystalline SiO$_2$[s] than with amorphous SiO$_2$[s]. \citet{dyrek_so2_2023} found a more broad cloud feature in the same wavelength range. Their retrieval model considers well-mixed cloud particles made from SiO[s], SiO$_2$[s], MgSiO$_3$[s], and C$_\mathrm{amorphous}$[s]. They predict considerably heterogeneous cloud particles. Both these findings agree well with our results. Within the transit spectrum of HATS-6b, WASP-39b, and the morning terminator of WASP-76b, all three non-mixed treatments show a clear peak in transit depth around 10~$\mu$m which is caused by homogeneous Mg$_2$SiO$_4$[s]. Well-mixed particles on the other hand only show a generally muted spectrum or a single broad feature. However, seeing a broad feature in the transit depth does not automatically prove that cloud particles are well-mixed. For WASP-107b (Sect.~\ref{sec:planets_wasp107b}), also non-mixed treatments show a broad cloud feature. Here, it is important to note that the shape of cloud particles also impacts the spectral features of cloud particles. One of the advantages of BAS and SSA is that they can more easily account for non-spherical particles. Only the Mie theory calculations needs to be adjusted which can be done using DHS, for example. Accounting for non-spherical particles is more difficult for well-mixed particles, since LLL and Bruggeman already assume that inclusions are spherical.

    In this study, we used a forward modelling approach which shows clear differences in the transit depth between the mixing treatments. Extracting information from observations, however, is more difficult. Cloud free evening terminators, like in the case of WASP-76b, can reduce the signal from clouds if limbs are not observed separately. For WASP-39b, the muting of molecular features by the cloud structure predicted by \citet{carone_wasp-39b_2023} is larger than the muting found in the observations of \citet{rustamkulov_early_2023}. While some mixing treatments do result in less muting than others, this alone cannot explain the differences between the forward modelling and the observations. Higher cloud mass fractions or the atmospheric extent of the clouds also impact the muting of molecular features, but it is difficult to disentangle their effects. This degeneracy can also affect retrieval models. In our analysis of WASP-107b (Sect.~\ref{sec:planets_wasp107b}), some mixing treatments overpredict the cloud contributions more than others. A retrieval run using our transmission spectra calculation could result in different cloud mass fractions, cloud particle radii, or vertical extend of the clouds depending on the mixing treatment selected. For a detailed analysis of how mixing treatments affect retrievals, a more detailed study is required which is beyond the scope of this work.

    \subsection{Iron-bearing species and carbon require precise modelling}
    \label{sec:dis_iron}

    In Sect.~\ref{sec:2comp}, we show that iron-bearing species (Fe[s], FeO[s], Fe$_2$O$_3$[s], Fe$_2$SiO$_4$[s], and FeS[s]) and carbon (C[s], C$_\mathrm{amorphous}$[s]) can significantly change the optical properties of cloud particles even at volume fractions of less than 1\%. The strong impact is caused by the large imaginary part of the refractive index $k$ of these species and the assumption of spherical particles (see Sect.~\ref{sec:2comp_emts}). Small amounts of iron-bearing species can lead to an effective refractive index closer to the resonance values of a spherical particle. This can lead to an increased absorption efficiency of mixed spherical particles compared to homogeneous spherical particles.

    \citet{min_absorption_2006} investigated how the shape of particles impact their optical properties. For irregularly shaped particles, they showed that species with high $n$ and $k$ values have a much larger absorption efficiency than spherical particles of the same material. Assuming spherical and homogeneous cloud particles therefore can underestimate the absorption efficiency of clouds. A detailed study on how EMTs and non-mixed treatments affect non-spherical particles is, while important, beyond the scope of this study.

    Our results show that precisely modelling the carbon and iron-bearing cloud particle materials is crucial to accurately predict the optical properties of clouds. This holds true for EMTs as well as non-mixed treatments and for spherical as well as non-spherical cloud particles. Observing exoplanets where it can be reasonably expected that iron-bearing species occur as cloud particle materials, like WASP-76b, might allow further insights into cloud particle properties in exoplanet atmospheres. In particular, the morning terminator of WASP-76b is expected to have iron-bearing clouds. Detailed observations of the muting of molecular features around 5~$\mu$m and the cloud features around 10~$\mu$m might help to differentiate cloud particle morphologies.

\section{Conclusion}
    \label{sec:conclusion}

    Cloud particles in exoplanet atmospheres might be considerably heterogeneous. We analysed how 21 common cloud particle materials and different assumptions on the mixing treatment of cloud particle materials impact cloud particle optical properties. In total 4 EMTs (Bruggeman, LLL, Maxwell-Garnett, Linear) and 3 non-mixed treatments (core-shell, BAS, SSA) were studied. The mixing treatments were used to calculate the transmission spectrum of the planets HATS-6b, WASP-39b, WASP-76b, and WASP-107b.

    Species with large refractive indices, like iron-bearing species (Fe[s], FeO[s], Fe$_2$O$_3$[s], Fe$_2$SiO$_4$[s], FeS[s]) or carbon (C[s], C$_\mathrm{amorphous}$[s]), can change the optical properties of cloud particles at volume fractions below 1\%. Therefore, it is crucial to accurately model such species. Other cloud materials like high temperature condensates (TiO$_2$[s], Al$_2$O$_3$[s], CaTiO$_3$[s]), magnesium-silicates (MgSiO$_3$[s], Mg$_2$SiO$_4$[s]) metal oxides (SiO[s], SiO$_2$[s], MgO[s], MnO[s]), and salts (KCl[s], NaCl[s]) also impact the transmission spectrum, but do so typically at volume fractions above 10\%. For sulphur-bearing species (ZnS[s], Na$_2$S[s], MnS[s]), the impact on the optical properties of heterogeneous cloud particles is different depending on the species.

    The mixing treatment of heterogeneous cloud particles impacts cloud spectral features as well as the muting of molecular features. All mixing treatments lead to a muting of molecular features, with LLL typically leading to the strongest muting. non-mixed treatments can retain the spectral features of individual cloud particle materials, whereas well-mixed theories typically exhibit broader features. For non-mixed particles, assuming a core-shell morphology, equally-sized homogeneous cloud particles (SSA), or equally-numbered homogeneous cloud particles (BAS) resulted in similar transit spectra. In particular SSA and core-shell were for most parts indistinguishable. For well-mixed particles, the differences in transit depth between Bruggeman and LLL can reach up to 500~ppm for WASP-39b and HATS-6b. Maxwell-Garnett and Linear are not suitable for cloud particles in exoplanet atmospheres.

    While we have shown the impact of cloud particle morphologies and composition on the transmission spectra of exoplanets, it is difficult to disentangle them observationally from cloud particle properties like the dust to gas-ratio. Here, observations of cloud spectral features around 10~$\mu$m in addition to the muting of molecular features at lower wavelengths might help to break these degeneracies. In particular, observations of planets were iron-bearing species can be expected to from clouds, like WASP-76b, could help to investigate the morphologies of heterogeneous cloud particles.

    \section{Data availability}
    The planetary parameters used for the transmission spectra calculation, information of the opacity data of the homogeneous materials, and additional figures are available in Zenodo at https://dx.doi.org/10.5281/zenodo.13373167.

\begin{acknowledgements}
    S.K., A.D.S, L.C., M.M., Ch.H, and L.D. acknowledge funding from the European Union H2020-MSCA-ITN-2019 under grant agreement no. 860470 (CHAMELEON). D.S. and D.A.L. acknowledge financial support and use of the computational facilities of the Space Research Institute of the Austrian Academy of Sciences.
\end{acknowledgements}

\bibliographystyle{aa} 
\bibliography{current_version} 

\clearpage

\begin{appendix}

    \section{Mie Theory}
    \label{app:sec:mietheory}
    Mie theory \citep{mie_beitrage_1908} describes the solution to the Maxwell equations for a sphere with an effective refractive index of $\epsilon_\mathrm{eff}$ in a medium with refractive index $\epsilon_\mathrm{m}$. For cloud particles the medium is vacuum and thus we have $\epsilon_\mathrm{m} = 1$. While Mie theory has a well defined solution, numerical implementations differ. Here we test three different implementations of calculate Mie-Theory:
    \begin{itemize}
        \item PyMieScatt \citep{sumlin_retrieving_2018}
        \item miepython \citep{wiscombe_mie_1979, prahl_miepython_2023}
        \item Miex \citep{wolf_mie_2004}
    \end{itemize}
    The solution of Mie theory includes an infinite sum over Riccati-Bessel functions. For a numerical solution a suitable maximum number of summation terms $N_\mathrm{max}$ has to be selected. According to \citet{wiscombe_mie_1979} the following $N_\mathrm{max}$ is a suitable choice:
    \begin{align}
        N_{\mathrm{max}} = x + 4.05 \sqrt[3]{x} + 2,
    \end{align}
    where $x$ is the size parameter. This $N_\mathrm{max}$ was used by both miepython and PyMieScatt. Miex calculates $N_\mathrm{max}$ based on \citet{loskutov_light_1971} which is given by:
    \begin{align}
       N_{\mathrm{max}} > | \epsilon_\mathrm{eff} |^2 * x + 28
    \end{align}
    However, it is worth mentioning that \citet{wiscombe_mie_1979} notes in their work that the scaling of the Mie scattering error with $N_{\mathrm{max}}$ dependants only slightly on the refractive index and mostly on the size parameter $x$. All three technique use the logarithmic derivative the Riccati-Bessel function $D_n$. The solution is derived following \cite{bohren_absorption_2008} and using (mostly) downward recursion starting from $N_{\mathrm{max}}$. PyMieScatt and Miex initialise the recursion with $D_\mathrm{max} = 0$. Miepython uses the Lentz method \citep{lentz_method_1973} to derive a more precise initial condition.

    We calculate the relative differences in extinction efficiency $Q_\mathrm{ext}$ and scattering efficiency $Q_\mathrm{sca}$ relative to Miex to test the difference in the three routines. The following materials are tested: Al$_2$O$_3$[s], Fe$_2$SiO$_4$[s], FeO[s], FeS[s], Mg$_2$SiO$_4$[s], MgO[s], SiO$_2$[s], SiO[s], and TiO$_2$[s]. We tested multiple particle sizes in the range of 0.001 $\mu$m up to 1 $\mu$m and found that $a = 0.1$ $\mu$m generally produced the largest differences. The relative difference between Miex and miepython is always below 10$^{-6}$ for the extinction efficiency and 10$^{-8}$ for the scattering efficiency. Between PyMieScatt and Miex, the relative difference reaches up to 31\% for the extinction efficiency and 1.3\% for the scattering efficiency. To investigate the impact of these differences on observations, we produced the transmission spectrum of HATS-6b \citep{hartman_hats-6b_2015, kiefer_under_2024} using all 3 routines. For this spectrum, we also considered the equilibrium chemistry from the kinetic cloud formation model. The relative differences in the transit depth between Miex, PyMieScatt and miepython are all below 10$^{-33}$. Since observational limits are in the order of 10$^{-5}$, there is no observable difference between all three routines.

    Overall, we found no significant differences between Miex and miepython. The differences between Miex and PyMieScatt do not impact the transmission spectrum. However, it cannot be ruled out that these differences can impact the thermal structure of an exoplanet atmospheres.

    \section{CaSiO$_3$ opacity treatment}
    \label{sec:app_casio3}

    \begin{figure}
        \centering
        \includegraphics[width=\hsize]{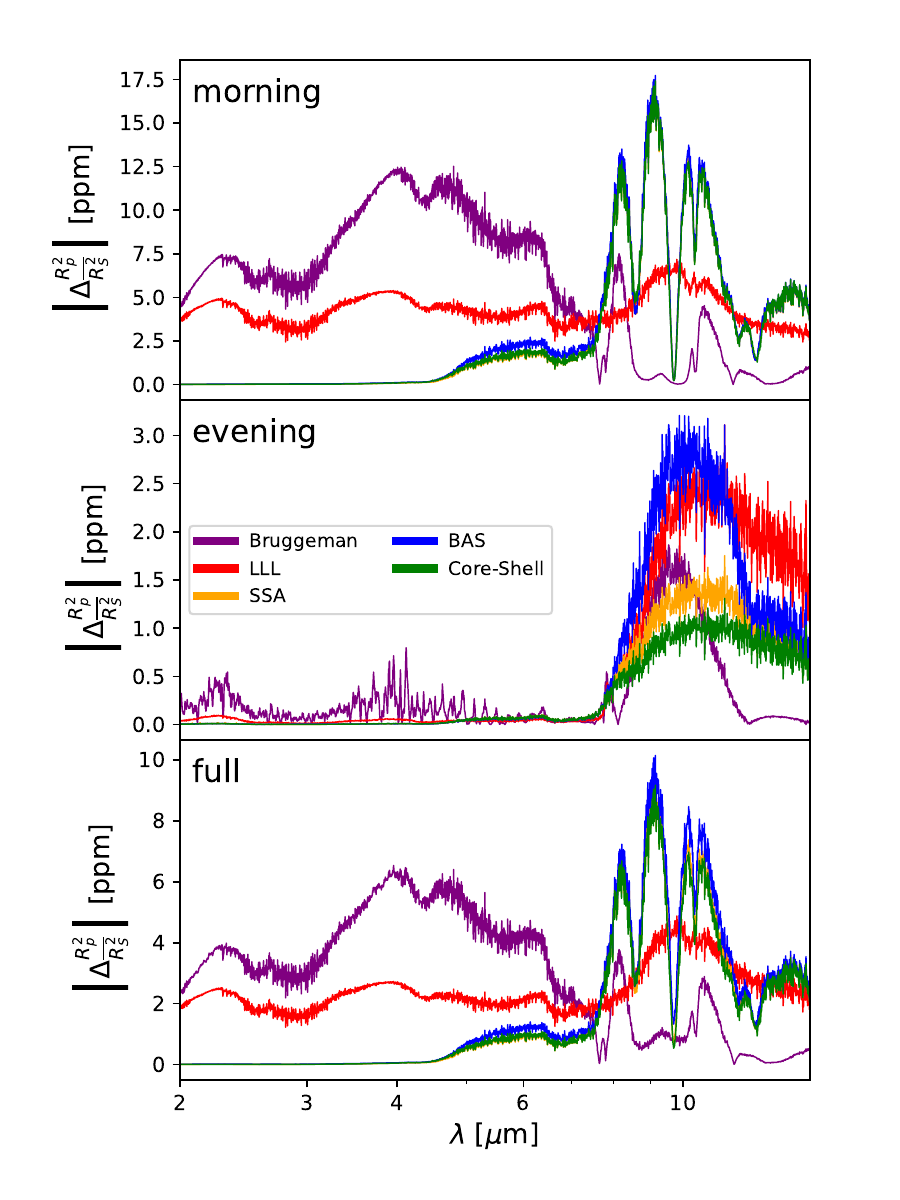}
        \caption{Absolute relative differences in the transmission spectra of WASP-76b between the approximation of CaSiO$_3$[s] refractive index with vacuum and MgSiO$_3$[s] values.}
        \label{fig:app_comp_vaccumassumption}
    \end{figure}

    Because no opacity data is available for CaSiO$_3$[s], the refractive index has to be approximated. One possibility is to use vacuum values. However, this will lead to an under-prediction of the cloud particle opacities. Another approach is to approximate the refractive index of CaSiO$_3$[s] with MgSiO$_3$[s]. While this approximation is likely closer to the real refractive index of CaSiO$_3$[s], it also introduces additional MgSiO$_3$[s] spectral features that might not be representative for CaSiO$_3$[s].

    From the four planets studied in Sect.~\ref{sec:planets}, WASP-76b has the highest CaSiO$_3$[s] volume fractions. We therefore calculate the transmission spectra of WASP-76b once using vacuum values and once using the refractive index values of MgSiO$_3$[s] for the refractive index of CaSiO$_3$[s]. The absolute relative difference between the transit depths can be seen in Fig.~\ref{fig:app_comp_vaccumassumption}. The two approximations lead to differences in transit depth of up to 17 ppm between 8~$\mu$m to 11~$\mu$m. This wavelength range corresponds to an increase in the imaginary part of the refractive index $k_\mathrm{MgSiO_3[s]}$ which is specific to MgSiO$_3$[s]. Between 2~$\mu$m to 4~$\mu$m $k_\mathrm{MgSiO_3[s]}$ is close to zero. In this wavelength range only the well-mixed particles show a difference in the transmission spectra. Overall, the differences in transit depth due to the two approximations for the CaSiO$_3$[s] opacity is much smaller than the general variations in transit depth.

    A compromise between the two approaches is using the wavelength-averaged values from MgSiO$_3$[s] instead of vacuum values. This results in refractive index values for CaSiO$_3$[s] that are closer to the values of MgSiO$_3$[s] without introducing additional spectral features. When taking the wavelength-average, it is important to exclude wavelengths where spectral features of MgSiO$_3$[s] occur because the refractive index values would otherwise be overestimated. The wavelength-averaged values for MgSiO$_3$[s] between $0.2~\mu$m and $9~\mu$m are $n_\mathrm{avg} = 1.544$ and $k_\mathrm{avg} = 5.5213 \times 10^{-4}$. We performed the same comparison as in Fig.~\ref{fig:app_comp_vaccumassumption} and found the differences in using the wavelength-averaged values of MgSiO$_3$[s] to using vacuum values is below 0.6 ppm in all cases.

\end{appendix}

\end{document}